# Light Types for Polynomial Time Computation in Lambda-Calculus [*]

Patrick Baillot
Laboratoire d'Informatique de Paris-Nord /CNRS
Université Paris-Nord, France
pb@lipn.univ-paris13.fr

Kazushige Terui
National Institute of Informatics
Tokyo, Japan
terui@nii.ac.jp

**Abstract**

*We propose a new type system for lambda-calculus ensuring that well-typed programs can be executed in polynomial time: Dual light affine logic (DLAL). DLAL has a simple type language with a linear and an intuitionistic type arrow, and one modality. It corresponds to a fragment of Light affine logic (LAL). We show that contrarily to LAL, DLAL ensures good properties on lambda-terms: subject reduction is satisfied and a well-typed term admits a polynomial bound on the reduction by any strategy. Finally we establish that as LAL, DLAL allows to represent all polytime functions.*

## 1 Introduction

Functional languages like ML assist the programmer with prevention of such errors as run-time type errors, thanks to automatic type inference. One could wish to extend this setting to verification of quantitative properties, such as time or space complexity bounds (see for instance [18]). We think that progresses on such issues can follow from advances in the topic of Implicit Computational Complexity, the field that studies calculi and languages with intrinsic complexity properties. In particular some lines of research have explored recursion-based approaches ([20, 7, 17, 8, 16]) and approaches based on linear logic to control the complexity of programs ([14, 19]).

Here we are interested in Light affine logic (LAL) ([2, 14]), a logical system designed from Linear logic and which characterizes polynomial time computation. By the Curry-Howard correspondence proofs in this logic can be used as programs. Some nice aspects of this system with respect to other approaches are the facts that it includes higher-order types as well as polymorphism. Moreover it naturally extends to a consistent naive set theory, in which one can reason about polynomial time concepts. In particular the provably total functions of that set theory are exactly the polynomial time functions ([14, 26]).

However the syntax of LAL is quite delicate, in particular because it has two modalities. Some term languages have been proposed (in particular in [25]) but programming is in general difficult. We think a better grasp would be given on this system if one could use as language plain lambda-calculus and then in a second phase have an automatic (or semi-automatic) LAL type inference performed. In case of success a well-typed program would have the guarantee that it can be executed in polynomial time.

This approach has been examined in [3, 4]. In particular it has been shown in [4] that type inference in propositional LAL is decidable. However some problems remain:

- First, to execute the well-typed program with the expected polynomial bound the lambda-term is not sufficient. One has to use the type derivation and extract a *light lambda term* (introduced in [25]) or a proof-net ([2]) that can be executed with the correct bound. In particular this means that if we use ordinary abstract machines for the evaluation we do not have any guarantee on the execution time.

- Second, even if type inference is decidable we do not have for the moment any efficient procedure. The difficulty actually comes from two points: the type derivation might need to specify some sharing of subterm; moreover the language of types is large (because there are two modalities) and this results in an important search space to explore.

To try to overcome these problems we propose here a new type system, that we call Dual light affine logic (DLAL). It corresponds to a simple fragment of LAL. It relies on the idea of replacing the ! modality by two notions of arrows: a linear one and an intuitionistic one. This is in the line of the works of Barber and Plotkin (Dual intuitionistic linear logic, [6]) and Benton ([9]). DLAL then offers the following advantages over LAL as a type system:

[*]Work partially supported by project GEOCAL ACI *Nouvelles interfaces des mathématiques*, project CRISS ACI *Sécurité informatique* (France) and Grant-in-Aid for Scientific Research, MEXT, Japan.

- Its language of types is 'smaller', in the sense that it corresponds to a strict subset of LAL types.

- DLAL keeps the same properties as LAL (P-completeness and polynomial bound on execution) but ensures the complexity bound on the lambda-term itself: if a term is typable one can extract the bound from the derivation, then forget about the type and execute the term using any strategy (and any abstract machine), with the guarantee that the reduction will terminate within the bound. This means that DLAL offers a system where the program part and the complexity specification part are really separate. The program part corresponds to the lambda-term and the complexity specification to the type.

- We think type inference should become easier, though this question still has to be explored. Indeed DLAL offers the following advantages: first there is no sharing in DLAL derivations; second, a large part of the difficulty of LAL type inference has to do with the fact that the types can use any sequence of the two modalities !, §, that is to say words over a binary alphabet. For this reason the type inference procedure of [4] used words constraints, which are hard to solve. By contrast Elementary affine logic (EAL) (corresponding to elementary complexity) has only one modality ! and its type inference can be performed using linear constraints, that is to say integer programming. The problem of EAL type inference has been shown decidable and studied in detail by Coppola *et al.* (see [11, 12]), starting from motivations in optimal reduction.

We believe DLAL should be easier to understand than LAL and could make this light logic approach accessible to a larger community. Moreover DLAL might open the way to a closer study of LAL types as well as of evaluation procedures for LAL-typed lambda-terms.

**Acknowledgements.** We are grateful to Paolo Coppola, Simone Martini and Ugo Dal Lago for their accurate reading and important suggestions.

## 2 Background on Light affine logic

**Notations**. Given a lambda-term $t$ we denote by $FV(t)$ the set of its free variables. Given a variable $x$ we denote by $no(x, t)$ the number of occurrences of $x$ in $t$. The notation $\longrightarrow$ will stand for $\beta$-reduction on lambda-terms. The size $|t|$ of a term is given by:

$$|x| = 1, \ |\lambda x.t| = |t| + 1, \ |(t \ u)| = |t| + |u| + 1.$$

### 2.1 Light affine logic

The formulas of (Intuitionistic) Light affine logic, LAL, are given by the following grammar:

$$A, B ::= \alpha \mid A \multimap B \mid !A \mid §A \mid \forall \alpha.A$$

We omit the connective $\otimes$ which is definable. We will write † instead of either ! or §.

Light affine logic is a logic for polynomial time computation in the proofs-as-programs approach to computing. It controls the number of reduction (or cut-elimination) steps of a proof-program using two ideas:
 (i) stratification,
 (ii) control on duplication.

Stratification means that the proof-program is divided into levels and that the execution preserves this organization. It is managed by the two modalities (also called *exponentials*) ! and §.

Duplication is controlled as in Linear logic: an argument can be duplicated only if it has undergone a !-rule (hence has a type of the form $!A$). What is specific to LAL with respect to Linear logic is the condition under which one can apply a !-rule to a proof-program: it should have at most one occurrence of free variable (rule (! i) of Figure 1).

We present the system as a natural deduction type-assignment system for lambda-calculus that we call NLAL: see Figure 1. We have:

- for ($\forall$ i): (*) $\alpha$ does not appear free in $\Gamma$.

- the (! i) rule can also be applied to a judgement of the form $; \vdash u : A$ ($u$ has no free variable).

This system uses the notion of *discharged formulas*, which are expressions of the form $[A]_\dagger$ with $\dagger = !$ or § (resp. !-discharged or §-discharged formula), where $A$ is a (proper) formula. Discharged formulas only appear on the l.h.s. of judgments and the only rules that can be applied to them are (! e), (§ e) and (Cntr). In particular note that one cannot apply the ($\multimap$ i) rule to a discharged formula. Discharged formulas are merely a technical artifact to handle the rules for modalities and contraction in a convenient way; in particular we do not use them in final typing judgments.

The notation $\Gamma, \Delta$ will be used for environments attributing formulas to variables. For environments of discharged formulas we use the following notation: if $\Gamma = x_1 : A_1, \ldots, x_n : A_n$ then $[\Gamma]_\dagger = x_1 : [A_1]_\dagger, \ldots, x_n : [A_n]_\dagger$. We also write $\dagger\Gamma = x_1 : \dagger A_1, \ldots, x_n : \dagger A_n$.

The sequent calculus presentation of LAL is perhaps better known, (we recall it in Appendix A) but natural deduction is more convenient for our purpose here. In the sequel we write $\Gamma \vdash_{LAL} t : A$ for a judgement derivable in NLAL.

The *depth* of a derivation $\mathcal{D}$ is the maximal number of (! $i$) and (§ $i$) rules in a branch of $\mathcal{D}$. We denote by $|\mathcal{D}|$ the *size* of $\mathcal{D}$ defined as its number of judgments.

$$\frac{}{x:A \vdash x:A} \text{ (Id)}$$

$$\frac{\Gamma, x:A \vdash t:B}{\Gamma \vdash \lambda x.t : A \multimap B} \text{ (}\multimap\text{ i)} \qquad \frac{\Gamma_1 \vdash t:A \multimap B \quad \Gamma_2 \vdash u:A}{\Gamma_1, \Gamma_2 \vdash (t\ u):B} \text{ (}\multimap\text{ e)}$$

$$\frac{\Gamma_1 \vdash t:A}{\Gamma_1, \Gamma_2 \vdash t:A} \text{ (Weak)} \qquad \frac{x_1:[A]_!, x_2:[A]_!, \Gamma \vdash t:B}{x:[A]_!, \Gamma \vdash t[x/x_1, x/x_2]:B} \text{ (Cntr)}$$

$$\frac{\Gamma, \Delta \vdash t:A}{[\Gamma]_!, [\Delta]_\S \vdash t:\S A} \text{ (}\S\text{ i)} \qquad \frac{\Gamma_1 \vdash u:\S A \quad \Gamma_2, x:[A]_\S \vdash t:B}{\Gamma_1, \Gamma_2 \vdash t[u/x]:B} \text{ (}\S\text{ e)}$$

$$\frac{x:B \vdash t:A}{x:[B]_! \vdash t:!A} \text{ (! i)} \qquad \frac{\Gamma_1 \vdash u:!A \quad \Gamma_2, x:[A]_! \vdash t:B}{\Gamma_1, \Gamma_2 \vdash t[u/x]:B} \text{ (! e)}$$

$$\frac{\Gamma \vdash t:A}{\Gamma \vdash t:\forall \alpha.A} \text{ (}\forall\text{ i) (*)} \qquad \frac{\Gamma \vdash t:\forall \alpha.A}{\Gamma \vdash t:A[B/\alpha]} \text{ (}\forall\text{ e)}$$

**Figure 1. Natural deduction for LAL**

Now, Light affine logic enjoys the following property:

**Theorem 1 ([14, 1])** *Given a NLAL proof $\mathcal{D}$ with depth $d$, its normal form $\mathcal{D}_0$ can be computed in $O(|\mathcal{D}|^{2^{d+1}})$ steps.*

This statement refers to reduction performed either on proof-nets ([14, 2]) or on light lambda terms ([25]). If the depth $d$ is fixed and the size of $\mathcal{D}$ might vary (for instance when applying a fixed term to binary integers) then the result can be computed in polynomial steps.

Moreover we have:

**Theorem 2 ([14, 2])** *If a function $f:\{0,1\}^\star \to \{0,1\}^\star$ is computable in polynomial time, then it is representable in LAL.*

## 2.2 LAL and beta-reduction

It was shown in [25] that light affine lambda-calculus admits polynomial strong normalization: the bound of theorem 1 holds on the length of *any* reduction sequence of light affine lambda-terms. However, this property is not true for LAL-typed plain lambda-terms and $\beta$-reduction: indeed [2] gives a family of LAL-typed terms (with a fixed depth) such that there exists a reduction sequence of exponential length. So the reduction of LAL-typed lambda-terms is not *strongly* poly-step (when counting the number of beta-reduction steps). Hence it is not strongly polytime, when counting the cost of the simulation of the reduction on a Turing machine.

We stress here with an example the fact that normalization of LAL-typed lambda-terms is not even *weakly* polytime: there exists a family of LAL-typed terms (with fixed depth) such that the computation of their normal form on a Turing machine (using any strategy) will take exponential space, hence exponential time.

First, observe that the following judgments are derivable:

$$y_i :!A \multimap !A \multimap !A \vdash_{LAL} \lambda x.y_i xx :!A \multimap !A$$

$$z :!A \vdash_{LAL} z :!A$$

From this it is easy to check that the following is derivable:
$y_1 :!A \multimap !A \multimap !A, \ldots, y_n :!A \multimap !A \multimap !A, z :!A \vdash (\lambda x.y_1 xx)(\cdots(\lambda x.y_n xx)z \cdots) :!A$

Using $(\S i)$, $(Cntr)$ and $(!e)$ we finally get:

$$y :!(!A \multimap !A \multimap !A), z :!!A \vdash (\lambda x.yxx)^n z : \S !A$$

Denote by $t_n$ the term $(\lambda x.yxx)^n z$ and by $u_n$ its normal form. We have $u_n = y\ u_{n-1} u_{n-1}$, so $|u_n| = O(2^n)$, whereas $|t_n| = O(n)$: the size of $u_n$ is exponential in the size of $t_n$. Hence computing $u_n$ from $t_n$ on a Turing machine will take at least exponential space (if the result is written on the tape as a lambda-term).

It should be noted though that even if $u_n$ is of exponential size, it nevertheless has a type derivation of size $O(n)$. To see this, note that we have $z : [A]_!, y :!A \multimap !A \multimap !A \vdash_{LAL} yzz :!A$. Now make $n$ copies of it and compose them by $(!\ e)$; each time $(!\ e)$ is applied, the term size is doubled. Finally, by applying $(!\ e)$, $(\S\ i)$, $(Cntr)$ and $(!\ e)$ as before, we obtain a linear size derivation for $y :!(!A \multimap !A \multimap !A), z :!!A \vdash_{LAL} u_n : \S !A$.

## 2.3 Discussion

The counter-example of the previous section illustrates a mismatch between lambda-calculus and Light affine logic. It can be ascribed to the fact that the $(!\ e)$ rule on lambda-calculus not only introduces sharing but also causes duplication. As Asperti neatly points out ([1]), "while every datum of type $!A$ is eventually sharable, not all of them are actually duplicable." The above $yzz$ gives a typical example. While it is of type $!A$ and thus sharable, it should not be duplicable, as it contains more than one free variable occurrence. The (! e) rule on lambda-calculus, however, neglects this delicate distinction, and actually causes duplication.

Light affine lambda-calculus ($\lambda$LA) remedies this by carefully designing the syntax so that the (! e) rule allows sharing but not duplication. As a result, it offers the properties of subject-reduction with respect to LAL and polynomial strong normalization ([25]). However it is not as simple as lambda-calculus; in particular it includes new constructions $!(.)$, $\S(.)$ and $\text{let}(.)$ be $\dagger\, x\, \text{in}\,(.)$ corresponding to the management of boxes and contractions in proof-nets.

The solution we propose here is more drastic: we simply do not allow the (! e) rule to be applied to a term of type $!A$. This is achieved by removing judgments of the form $\Gamma \vdash t :!A$. As a consequence, we also remove types of the form $A \multimap !B$. Bang $!$ is used only in the form $!A \multimap B$, which we consider as a primitive connective $A \Rightarrow B$. Note that it hardly causes a loss of expressiveness in practice, since linear logic as decomposition of intuitionistic logic does not use types of the form $A \multimap !B$.

## 3  Dual light affine logic (DLAL)

The system we propose does not use the $!$ connective but distinguishes two kinds of function spaces (linear and non-linear). This approach is analogous to that of Dual intuitionistic linear logic of Barber and Plotkin ([6]), or the system of Benton ([9]), which correspond to Intuitionistic linear logic. Thus we call our system Dual light affine logic (DLAL). We will see that it corresponds in fact to a well-behaved fragment of LAL.

The language $\mathcal{L}_{DLAL}$ of DLAL types is given by:

$$A, B ::= \alpha \mid A \multimap B \mid A \Rightarrow B \mid \S A \mid \forall \alpha.A$$

There is an unsurprising translation $(.)^*$ from DLAL to LAL given by:

- $(A \Rightarrow B)^* = !A^* \multimap B^*$,
- $(.)^*$ commutes to the other connectives.

Let $\mathcal{L}_{DLAL\star}$ denote the image of $\mathcal{L}_{DLAL}$ by $(.)^*$.

For DLAL typing we will handle judgements of the form $\Gamma; \Delta \vdash t : C$. The intended meaning is that variables in $\Delta$ are (affine) linear, that is to say that they have at most one occurrence in the term, while variables in $\Gamma$ are non-linear. We give the typing rules as a natural deduction system that we call NDLAL: see Figure 2. There is only one kind of discharged formulas, $[A]_\S$, which as in the case of NLAL are not used in final typing judgments. We have:

- (*) $\alpha$ does not appear free in $\Gamma_1, \Delta_1$.

- in the ($\Rightarrow$ e) rule the r.h.s. premise can also be of the form $; \vdash u : A$ ($u$ has no free variable).

An alternative sequent calculus presentation is given in Appendix B.

In the rest of the paper we will write $\Gamma; \Delta \vdash_{DLAL} t : A$ for a judgement derivable in NDLAL.

**Remark 3** *In fact one could give an alternative presentation of NLAL without discharged formulas: for that one would replace the rules ($\S$ i), ($\S$ e) by a single rule with several premises (in the style of [10]). The properties of the system would be the same; we adopted the present formulation because it is slightly more convenient to prove the properties in the next sections.*

Observe that the contraction rule (Cntr) is used only on variables on the l.h.s. of the semi-column. It is then straightforward to check the following statement:

**Lemma 4** *If $\Gamma; \Delta \vdash_{DLAL} t : A$ then the set $FV(t)$ is included in the variables of $\Gamma \cup \Delta$, and if $x \in \Delta$ then we have $no(x, t) \leqslant 1$.*

We can make the following remarks on NDLAL rules:

- Initially the variables are linear (rule (Id)); to convert a linear variable into a non-linear one we have to use the ($\S$ i) rule. Note that it adds a $\S$ to the type of the result and that the variables that remain linear (the $x_i$) get a discharged type.

- the ($\multimap$ i) (resp. ($\Rightarrow$ i)) rule corresponds to abstraction on a linear variable (resp. non-linear variable);

- observe ($\Rightarrow$ e): a term of type $A \Rightarrow B$ can only be applied to a term $u$ with at most one occurrence of free variable.

Note that the only rules which correspond to substitutions in the term are (Cntr) and ($\S$ e): in (Cntr) only a variable is substituted and in ($\S$ e) substitution is performed on a linear variable. Combined with Lemma 4 this ensures the following important property:

**Proposition 5** *If a derivation $\mathcal{D}$ has conclusion $\Gamma; \Delta \vdash_{DLAL} t : A$ then we have $|t| \leq |\mathcal{D}|$.*

This Proposition shows that the mismatch between lambda-calculus and LAL illustrated in the previous section is resolved with DLAL.

One can observe that the rules of DLAL are obtained from the rules of LAL and the $(.)^*$ translation, and it follows that:

**Proposition 6** *Given a lambda-term $t$, if $\Gamma; \Delta \vdash_{DLAL} t : A$ then $[\Gamma^*]_!, \Delta^* \vdash_{LAL} t : A^*$.*

$$\frac{}{;x:A \vdash x:A}\ (\text{Id})$$

$$\frac{\Gamma_1;\Delta_1,x:A \vdash t:B}{\Gamma_1;\Delta_1 \vdash \lambda x.t:A \multimap B}\ (\multimap \text{i}) \qquad \frac{\Gamma_1;\Delta_1 \vdash t:A \multimap B \quad \Gamma_2;\Delta_2 \vdash u:A}{\Gamma_1,\Gamma_2;\Delta_1,\Delta_2 \vdash (t\ u):B}\ (\multimap \text{e})$$

$$\frac{\Gamma_1,x:A;\Delta_1 \vdash t:B}{\Gamma_1;\Delta_1 \vdash \lambda x.t:A \Rightarrow B}\ (\Rightarrow \text{i}) \qquad \frac{\Gamma_1;\Delta_1 \vdash t:A \Rightarrow B \quad ;z:C \vdash u:A}{\Gamma_1,z:C;\Delta_1 \vdash (t\ u):B}\ (\Rightarrow \text{e})$$

$$\frac{\Gamma_1;\Delta_1 \vdash t:A}{\Gamma_1,\Gamma_2;\Delta_1,\Delta_2 \vdash t:A}\ (\text{Weak}) \qquad \frac{x_1:A,x_2:A,\Gamma_1;\Delta_1 \vdash t:B}{x:A,\Gamma_1;\Delta_1 \vdash t[x/x_1,x/x_2]:B}\ (\text{Cntr})$$

$$\frac{;\Gamma,x_1:B_1,\ldots,x_n:B_n \vdash t:A}{\Gamma;x_1:[B_1]_\S,\ldots,x_n:[B_n]_\S \vdash t:\S A}\ (\S\ \text{i}) \qquad \frac{\Gamma_1;\Delta_1 \vdash u:\S A \quad \Gamma_2;x:[A]_\S,\Delta_2 \vdash t:B}{\Gamma_1,\Gamma_2;\Delta_1,\Delta_2 \vdash t[u/x]:B}\ (\S\ \text{e})$$

$$\frac{\Gamma_1;\Delta_1 \vdash t:A}{\Gamma_1;\Delta_1 \vdash t:\forall\alpha.A}\ (\forall\ \text{i})\ (*) \qquad \frac{\Gamma_1;\Delta_1 \vdash t:\forall\alpha.A}{\Gamma_1;\Delta_1 \vdash t:A[B/\alpha]}\ (\forall\ \text{e})$$

**Figure 2. Natural deduction for DLAL**

The data types of LAL can be directly adapted to DLAL. For instance we had as type for tally integers in LAL $N^{LAL} = \forall\alpha.!(\alpha \multimap \alpha) \multimap \S(\alpha \multimap \alpha)$, and in DLAL:

$$\begin{aligned} N &= \forall\alpha.(\alpha \multimap \alpha) \Rightarrow \S(\alpha \multimap \alpha) \\ W &= \forall\alpha.(\alpha \multimap \alpha) \Rightarrow (\alpha \multimap \alpha) \Rightarrow \S(\alpha \multimap \alpha) \end{aligned}$$

The type $W$ is a type for binary words. The inhabitants of type $N$ are the familiar Church integers:

$\underline{n} = \lambda f.\lambda x.(f\ (f\ldots(f\ x)\ldots))$

with $n$ occurrences of $f$. The following terms for addition and multiplication on Church integers are typable in DLAL:

$$\begin{aligned} add &= \lambda n.\lambda m.\lambda f.\lambda x.(n\ f\ (m\ f\ x)) : N \multimap N \multimap N \\ mult &= \lambda n.\lambda m.(m\ \lambda k.\lambda f.\lambda x.(n\ f\ (k\ f\ x)))\ \underline{0} \\ mult &: N \Rightarrow N \multimap \S N \end{aligned}$$

Finally, we have a partial converse to Proposition 6:

**Proposition 7** *If the following conditions hold:*

- *$t$ is in normal form,*
- *the judgment $[\Gamma']_!, \Delta' \vdash_{LAL} t : A'$ can be derived using ($\forall$ e) only with instantiation on $\mathcal{L}_{DLAL\star}$ formulas,*

*then the judgment $\Gamma;\Delta \vdash_{DLAL} t : A$ with $\Gamma^* = \Gamma'$, $\Delta^* = \Delta'$, $A^* = A'$ is derivable.*

To prove this Proposition we use the sequent calculus presentation of DLAL. The proof is given in appendix C.

## 4 Properties of DLAL

### 4.1 Subject reduction

In this section, we will establish the subject reduction property for DLAL. It should be stressed that subject reduction is by no means a trivial property in the current setting, because lambda-calculus does not have any constructs corresponding to modalities of light logics; as a matter of fact, LAL as a type assignment system for lambda-calculus (Figure 1) does not satisfy the subject reduction property. For this reason, we will give a rather detailed argument here. Throughout this section, by $\Gamma;\Delta \vdash t : A$ we will mean $\Gamma;\Delta \vdash_{DLAL} t : A$. We will also use notation $\Gamma;\Delta \vdash^n t : A$ when $\Gamma;\Delta \vdash t : A$ has a derivation of size at most $n$.

**Lemma 8 (Substitution)**

*(1) If $\Gamma;\Delta \vdash^n t : A$, then $\Gamma[B/\alpha];\Delta[B/\alpha] \vdash^n t : A[B/\alpha]$ for every $B$.*

*(2) If $\Gamma_1;\Delta_1 \vdash^n u : A$ and $\Gamma_2;x:A,\Delta_2 \vdash^m t : B$, then $\Gamma_1,\Gamma_2;\Delta_1,\Delta_2 \vdash^{n+m} t[u/x] : B$.*

*(3) If $;\Gamma_1,\Delta_1 \vdash^n u : A$ and $\Gamma_2;x:[A]_\S,\Delta_2 \vdash^m t : B$, then $\Gamma_1,\Gamma_2;[\Delta_1]_\S,\Delta_2 \vdash^{n+m} t[u/x] : B$.*

*(4) If $;z:C \vdash u : A$ and $x_1 : A, \ldots, x_n : A, \Gamma;\Delta \vdash t : B$, then $z:C,\Gamma;\Delta \vdash t[u/x_1,\ldots,u/x_n] : B$.*

A proof is given in Appendix D.1.

**Definition 1** *The l.h.s. premises of ($\multimap$ e), ($\Rightarrow$ e) and ($\S$ e) as well as the unique premise of ($\forall$ e) are called* major *premises. A DLAL derivation is $\forall\S$-normal if*

- *no conclusion of a ($\forall$ i) rule is the premise of a ($\forall$e) rule;*
- *no conclusion of a ($\S$ i) rule is the major premise of a ($\S$ e) rule;*
- *no conclusion of (Weak), (Cntr) and ($\S$ e) is the major premise of elimination rules: ($\multimap$ e), ($\Rightarrow$ e), ($\S$ e), ($\forall$ e).*

**Lemma 9 (∀§-Normalization)** *If $\Gamma; \Delta \vdash t : A$ has a derivation, then it also has a ∀§-normal derivation.*

This lemma can be proved by employing Substitution Lemma (1) and (3) as well as permutability of (Weak), (Cntr) and (§ e) over the elimination rules (see Appendix D.2).

**Lemma 10 (Abstraction Property)** *Let $\Gamma; \Delta \vdash \lambda x.t : A$ be derivable with a ∀§-normal derivation $\mathcal{D}$. Suppose that the last rule (r) of $\mathcal{D}$ is neither (Weak), (Cntr) nor (§ e). Then, (r) is an introduction rule corresponding to the outermost connective of $A$.*

*Proof.* By induction on $\mathcal{D}$. First, (r) cannot be (∀ e); if it were, then $\mathcal{D}$ would be of the form

$$\frac{\frac{\vdots}{\Gamma; \Delta \vdash \lambda x.t : \forall \alpha.B} \text{ (r')}}{\Gamma; \Delta \vdash \lambda x.t : B[C/\alpha]} \text{ (∀ e)}$$

Since $\mathcal{D}$ is ∀§-normal, (r') is neither (weak), (cntr) nor (§ e). Hence by the induction hypothesis, (r') must be (∀ i), but that is impossible.

Second, (r) cannot be (⊸ e), (⇒ e) nor (Id), since the subject $\lambda x.t$ does not match the subjects of these rules. The only possibility is therefore an introduction rule corresponding to the outermost connective of $A$.

As a direct consequence, we have:

**Lemma 11 (Paragraph Property)** *Let $\mathcal{D}$ be a ∀§-normal derivation. If $\mathcal{D}$ contains an application of (§ e):*

$$\frac{\Gamma_1; \Delta_1 \vdash u : §A \quad \Gamma_2; x : [A]_§, \Delta_2 \vdash t : B}{\Gamma_1, \Gamma_2; \Delta_1, \Delta_2 \vdash t[u/x] : B} \text{ (§ e)}$$

*then $u$ is not of the form $\lambda x.v$.*

*Proof.* Since $\mathcal{D}$ is assumed to be ∀§-normal, the last rule used for deriving the l.h.s. premise is neither (Weak), (Cntr) nor (§ e). Hence by the previous lemma, if $u$ is of the form $\lambda x.v$, the last rule must be (§ i), which contradicts the ∀§-normality of $\mathcal{D}$.

**Theorem 12 (Subject Reduction)** *If $\Gamma; \Delta \vdash t_0 : A$ is derivable and $t_0 \longrightarrow t_1$, then $\Gamma; \Delta \vdash t_1 : A$ is derivable.*

*Proof.* By ∀§-Normalization Lemma, there is a ∀§-normal derivation $\mathcal{D}$ of $\Gamma; \Delta \vdash t : A$. The proof is carried out by induction on $\mathcal{D}$.

(Case 1) The last rule of $\mathcal{D}$ is (⊸ e):

$$\frac{\begin{array}{c}\vdots \mathcal{D}_1 \\ \Gamma_1; \Delta_1 \vdash t : A \multimap B\end{array} \quad \begin{array}{c}\vdots \mathcal{D}_2 \\ \Gamma_2; \Delta_2 \vdash u : A\end{array}}{\Gamma_1, \Gamma_2; \Delta_1, \Delta_2 \vdash (t\ u) : B} \text{ (⊸ e)}$$

If the redex is inside $t$ or $u$, then the statement of the theorem follows from the induction hypothesis. If $(t\ u)$ itself is the redex, then $t$ must be of the form $\lambda x.v$. By Abstraction Property Lemma, the last rule of $\mathcal{D}_1$ is (⊸ i), hence we have $\Gamma_1; x : A, \Delta_1 \vdash v : B$. By Substitution Lemma (2), we have $\Gamma_1, \Gamma_2; \Delta_1, \Delta_2 \vdash v[u/x] : B$ as required.

(Case 2) The last rule of $\mathcal{D}$ is (⇒ e): Similar to (Case 1), except that Substitution Lemma (4) is used instead of (2).

(Case 3) The last rule is (§ e):

$$\frac{\Gamma_1; \Delta_1 \vdash u : §A \quad \Gamma_2; x : [A]_§, \Delta_2 \vdash t : B}{\Gamma_1, \Gamma_2; \Delta_1, \Delta_2 \vdash t[u/x] : B} \text{ (§ e)}$$

By Paragraph Property Lemma, $u$ is not an abstraction. Therefore, no new redex is created by substituting $u$ for $x$ in $t$. Thus each redex in $t[u/x]$ has a counterpart in $t$ or $u$, and we can therefore apply the induction hypothesis to obtain the desired result.

The other cases are straightforward.

### 4.2 Normalization

The *depth* of a DLAL derivation $\mathcal{D}$ is the maximal number of premises of (§ i) and r.h.s. premises of (⇒ e) in a branch of $\mathcal{D}$. DLAL types ensure the following strong normalization property:

**Theorem 13 (Polynomial time strong normalization)**
*Let $t$ be a lambda-term which has a typing derivation $\mathcal{D}$ of depth $d$ in DLAL. Then $t$ reduces to the normal form $u$ in at most $|t|^{2^d}$ reduction steps and in time $O(|t|^{2^{d+2}})$ on a Turing machine. This result holds independently of which reduction strategy we take.*

Here we prove a weaker form of the above theorem, namely we prove that there *exists* a reduction sequence from $t$ to $u$ which is of length at most $|t|^{2^d}$ and which requires time $O(|t|^{2^{d+2}})$ to execute. Although the result is weaker, it may be helpful for getting an idea of polynomial time normalization without recourse to LAL. Theorem 13 itself can be proved by showing that any beta reduction sequence for a DLAL typable lambda term can be simulated by a longer $\lambda$LA reduction sequence (see Appendix E).

**Definition 2** *A* stratified term *is a term with each abstraction symbol $\lambda$ annotated by a natural number $d$ (called its* depth*) and also possibly by symbol !.*

Thus an abstraction looks like $\lambda^d x.t$ or $\lambda^{d!} x.t$. In the following, $\lambda^{d!} x.t$ stands for either $\lambda^d x.t$ or $\lambda^{d!} x.t$. When $t$ is a stratified term, $t[+1]$ denotes $t$ with the depths of all abstraction subterms increased by 1. The type assignment rules for stratified terms are obtained by modifying the rules ($\multimap$ i), ($\Rightarrow$ i), ($\Rightarrow$ e), ($\S$ i) of DLAL as follows:

$$\frac{\Gamma_1; \Delta_1, x:A \vdash t:B}{\Gamma_1;\Delta_1 \vdash \lambda^0 x.t : A \multimap B} \ (\multimap \text{ i})$$

$$\frac{\Gamma_1, x:A; \Delta_1 \vdash t:B}{\Gamma_1;\Delta_1 \vdash \lambda^{0!} x.t : A \Rightarrow B} \ (\Rightarrow \text{ i})$$

$$\frac{\Gamma_1;\Delta_1 \vdash t:A \Rightarrow B \quad ;z:C \vdash u:A}{\Gamma_1, z:C; \Delta_1 \vdash (t\ u[+1]) : B} \ (\Rightarrow \text{ e})$$

$$\frac{; \Delta_1, \Delta_2 \vdash t:A}{\Delta_1; [\Delta_2]_\S \vdash t[+1] : \S A} \ (\S \text{ i})$$

A redex is *at depth* $d$ when its main abstraction is at depth $d$. The *depth* of a term $t$ is the maximal depth of all abstractions in it. We write $t \xrightarrow{d}{}^* u$ when there is a reduction sequence from $t$ to $u$ which consists of reductions of redexes at depth $d$.

**Lemma 14** *Given a DLAL derivation of $\Gamma; \Delta \vdash t : A$ of depth $d$, $t$ can be decorated as a stratified term $t'$ of depth $d$ such that $\Gamma; \Delta \vdash t' : A$.*

It is not hard to see that $\forall\S$-Normalization Lemma, Abstraction Property Lemma, Paragraph Property Lemma and Subject Reduction Theorem hold for stratified terms as well.

The following three lemmas are all concerned with typable stratified terms.

**Lemma 15** *Reducing a redex at depth $d$ does not create a new redex at depth less than $d$.*

*Proof.* We prove that there is no typable stratified term which contains a subterm of the form

(1) $(\lambda^{dl}x.t)(\lambda^{el}y.u)$ with $e < d$;

(2) $\lambda^{dl}x.\lambda^{el}y.t$ with $e < d$.

The lemma easily follows from this, because a lower depth redex is created only by reducing (1) or a redex of the form: $(\lambda^{dl}x.\lambda^{el}y.t)uv$ with $e < d$.

The above claim is proved by induction on the size of $\forall\S$-normal derivation $\mathcal{D}$.

(Case 1) The last inference is ($\multimap$ i): Since the rule ($\multimap$ i) always introduces an abstraction at depth 0, a term of the form (2) is never produced.

(Case 2) The last inference is ($\multimap$ e):

$$\frac{\Gamma_1;\Delta_1 \vdash t:A \multimap B \quad \Gamma_2;\Delta_2 \vdash u:A}{\Gamma_1, \Gamma_2; \Delta_1, \Delta_2 \vdash (t\ u) : B} \ (\multimap \text{ e})$$

If $t$ is an abstraction, then the last inference to derive $\Gamma_1; \Delta_1 \vdash t : A \multimap B$ is not (Weak), (Cntr) nor ($\S$ e), since $\mathcal{D}$ is $\forall\S$-normal. By Abstraction Property Lemma, the last inference should be ($\multimap$ i) and $t$ should be of the form $\lambda^0 x.t'$. Hence a term of the form (1) is never produced.

(Case 3) The last inference is ($\S$ e):

$$\frac{\Gamma_1;\Delta_1 \vdash u:\S A \quad \Gamma_2; x:[A]_\S, \Delta_2 \vdash t:B}{\Gamma_1, \Gamma_2; \Delta_1, \Delta_2 \vdash t[u/x] : B} \ (\S \text{ e})$$

By Paragraph Lemma, $u$ is not an abstraction. Hence a subterm of the form (1) or (2) is never produced by the substitution $t[u/x]$.

**Lemma 16** *If $t \xrightarrow{d}{}^* u$, then the length of the reduction sequence is bounded by $|t|$.*

*Proof.* Observe that:

- If a typable stratified term $t$ contains $(\lambda^d x.u)v$, then $no(x, u) \leq 1$ (see Lemma 4).

- If a typable stratified term $t$ contains $(\lambda^{d!} x.u)v$, then $v$ does not contain any abstractions at depth $d$.

Hence a reduction at depth $d$ strictly decreases the number of abstractions at depth $d$, that is obviously bounded by $|t|$.

**Lemma 17** *If $|t| \geq 2$ and $t \xrightarrow{d}{}^* u$, then $|u|$ is bounded by $|t|(|t| - 1)$.*

*Proof (sketch).* Observe that:

- Reducing a linear redex $(\lambda^d x.v_1)v_2$ does not increase the size.

- The number of *bound variables at depth $d$* (i.e. those bound by $\lambda^{dl}$) in $t$ is at most $|t| - 1$ (trivial).

- The above number does not increase by a reduction $C[(\lambda^{d!} x.v_1)v_2] \longrightarrow C[v_1[v_2/x]]$, because $v_2$ contains at most one free variable (which is possibly bound by another $\lambda^{d!}$ in the context $C$), and all other variables in $v_2$ are bound at a depth strictly greater than $d$.

Now, we can also note that:

- A reduction $(\lambda^{d!} x.v_1)v_2 \longrightarrow v_1[v_2/x]$ *produces* $n$ copies of $v_2$ and *consumes* $n$ occurrences of the bound variable $x$ at depth $d$ instead.

- It is possible that the above $v_2$ is substituted into a subterm $v_3[x]$ which is to be duplicated later. Note that such a duplicable subterm $v_3[x]$ may have at most one occurrence of a free variable $x$ due to the restriction on the rule ($\Rightarrow$ e). Therefore, when another reduction applies to a redex of the form $(\lambda^{d!} y.v')v_3[v_2]$, it produces $m$ copies of $v_2$, consuming $m$ occurrences of the bound variable $y$ at depth $d$ at the same time.

As a result, every subterm of $t$ which is to be duplicated during the reductions at depth $d$ gives rise to at most $|t| - 1$ copies in $u$.

Therefore, we conclude that the size of $u$ is bounded by $|t|(|t| - 1)$.

**Theorem 18 (Polynomial time weak normalization)** *Let $t$ be a lambda-term which has a typing derivation $\mathcal{D}$ of depth $d$ in DLAL. Then $t$ can be normalized within $|t|^{2^d}$ reduction steps, and within time $O(|t|^{2^{d+2}})$ on a Turing machine.*

*Proof.* By Lemma 14, $t$ can be decorated as a stratified term $t'$ of depth $d$. By Lemma 15, normalization can be done *by levels*. Namely, there is a reduction sequence of the form
$$t' \equiv t_0 \xrightarrow{0}{}^* t_1 \xrightarrow{1}{}^* \cdots t_d \xrightarrow{d}{}^* u$$
with $u$ normal. Without loss of generality, we may assume that $|t_i| \geq 2$ for $0 \leq i < d$. The length of the reduction sequence above is bounded by $|t_0| + |t_1| + \cdots + |t_d|$ by Lemma 16. Hence it is sufficient to show that
$$|t_0| + |t_1| + \cdots + |t_d| \leq |t|^{2^d}.$$

The proof is by induction on $d$. Since it is trivial when $d = 0$, let us assume $d > 0$. Then we have:

$$\begin{aligned}
\sum_{i=0}^{d} |t_i| &\leq |t|^{2^{d-1}} + |t_d| \text{ (by the induction hypothesis)} \\
&\leq |t|^{2^{d-1}} + |t_{d-1}|(|t_{d-1}| - 1) \text{ (by Lemma 17)} \\
&\leq |t|^{2^{d-1}} + |t|^{2^{d-1}}(|t|^{2^{d-1}} - 1) \\
&\quad \text{(by the induction hypothesis)} \\
&= |t|^{2^d}.
\end{aligned}$$

It is readily seen that the number $|t|^{2^d}$ also bounds the size of every term occurring in the above reduction sequence. Since a beta reduction step $t \longrightarrow u$ costs time $O(|t|^2)$ on a Turing machine, the overall time required for normalization is $|t|^{2^d} \cdot O(|t|^{2^d \cdot 2}) \leq O(|t|^{2^{d+2}})$.

## 4.3 Expressiveness

We will show that polynomial time Turing machines can be simulated in DLAL by adapting the proof given for LAL in [2]. The key point is that of coercions for type $N$.

### 4.3.1 Coercions

Coercions will allow us under certain conditions to turn a non-linear variable of integer type $N$ into a linear variable, and a linear variable of type $\S N$ into a linear variable of type $N$. We express coercions on the type $N$ as rules derivable in NDLAL:

$$\frac{n : N; \Delta \vdash t : A}{; m : N, \S\Delta \vdash C_1[t] : \S A} \text{ (coerc1)}$$

$$\frac{\Gamma; n : \S N, \Delta \vdash t : A}{\Gamma; m : N, \Delta \vdash C_2[t] : A} \text{ (coerc2)}$$

where $C_1[.]$ and $C_2[.]$ are contexts, which contain as free variables some variables of the environments:

$$\begin{aligned}
C_1[x] &= (m(\lambda g.\lambda p.(g \ (succ \ p))) \ \lambda n.x)\underline{0} \\
C_2[x] &= (\lambda n.x)(m \ succ \ \underline{0})
\end{aligned}$$

and $succ$ is the usual term for successor. Observe that in the conclusion of (coerc2) the context and the type of the term are not changed, while they are in (coerc1). Note also that in the premise of (coerc1) the variable $n$ is the only non-linear variable of the context.

**Lemma 19** *For $i = 1, 2$ we have: for any Church integer $\underline{k}$ and term $t$ the term $C_i[t][\underline{k}/m]$ reduces to $t[\underline{k}/n]$. Hence $\lambda m.C_i[t]$ is extensionally equivalent to $\lambda n.t$.*

For instance, $C_1[t][\underline{2}/m]$ reduces to $t[\underline{2}/n]$ as follows:

$$\begin{aligned}
C_1[t][\underline{2}/m] &\longrightarrow ((\lambda g.\lambda p.g \ (succ \ p))^2 \lambda n.t) \ \underline{0} \\
&\longrightarrow^* ((\lambda g.\lambda p.g \ (succ \ p))(\lambda p.t[succ \ p/n])) \ \underline{0} \\
&\longrightarrow^* (\lambda p.(t[succ \ succ \ p/n])) \ \underline{0} \\
&\longrightarrow t[succ \ succ \ \underline{0}/n] \\
&\longrightarrow^* t[\underline{2}/n].
\end{aligned}$$

### 4.3.2 Encoding some polynomials

For the simulation we need to encode polynomials on the type $N$. To keep things short and as it is sufficient for the Turing machines we will content ourselves with the family of polynomials of the form:

$$P[X] = aX^d + b, \quad \text{with } a, b \in \mathbb{N} \text{ and } d = 2^k.$$

We will use the technique of [22]. Recall from section 3 that we have:

$$add : N \multimap N \multimap N \qquad mult : N \Rightarrow N \multimap \S N.$$

Using successively the rules (coerc1), (coerc2), ($\S$ i), (Cntr), (coerc1) and ($\multimap$ i), we get from the typing judgment of $mult$ a judgment $;\vdash square : N \multimap \S^4 N$ (Figure 3). The term $square$ computes the squaring function.

By composing $square$ $k$ times using the $\S$ rules we get a term $u$ representing the function $x \longrightarrow x^{2^k}$ with type $N \multimap \S^{4k} N$.

We can derive for multiplication, using (coerc 1) and the rules for $\S$, a term $mult_p : \S^p N \multimap \S^{p+1} N \multimap \S^{p+2} N$ and for addition a term $add_q : \S^q N \multimap \S^q N \multimap \S^q N$. The Church integers $\underline{a}$ and $\underline{b}$ representing $a$ and $b$ can be given

$$
\begin{array}{c}
\dfrac{n_1 : N; n_2 : N \vdash \mathit{mult}\ n_1 n_2 : \S N}{\ ; m_1 : N, n_2 : \S N \vdash C_1[\mathit{mult}\ n_1 n_2] : \S^2 N} \text{(coerc1)} \\
\dfrac{}{\ ; m_1 : N, m_2 : N \vdash C_2[C_1[\mathit{mult}\ n_1 n_2]] : \S^2 N} \text{(coerc2)} \\
\dfrac{}{m_1 : N, m_2 : N; \vdash C_2[C_1[\mathit{mult}\ n_1 n_2]] : \S^3 N} (\S\ i) \\
\dfrac{}{m : N; \vdash C_2[C_1[\mathit{mult}\ n_1 n_2]][m/m_1, m_2] : \S^3 N} \text{(Cntr)} \\
\dfrac{}{\ ; m : N \vdash t : \S^4 N} \text{(coerc1)} \\
\dfrac{}{\ ; \vdash \mathit{square} : N \multimap \S^4 N} (\multimap i)
\end{array}
$$

**Figure 3. Type derivation for the Squaring function**

types $\S^p N$ and $\S^q N$. Hence, assuming $k \geq 1$ and taking $p = 4k-1$, $q = 4k+1$ we finally get the following term representing the polynomial $P$:

$$t_P = \lambda n.(\mathit{add}_q(\mathit{mult}_p\ \underline{a}\ (u\ n)))\underline{b} : N \multimap \S^q N.$$

### 4.3.3 Simulation of Ptime Turing machines

The encoding of a Ptime Turing machine in LAL ([2]) can be described in two parts: (i) the quantitative part: encoding the polynomial, (ii) the qualitative part: defining a function of type $\mathit{config} \multimap \mathit{config}$ where $\mathit{config}$ is the type of configurations, which simulates an execution step of the machine.

The whole encoding then exploits these two parts to iterate a suitable number of times the step function on the initial configuration.

One can check on the LAL derivations of [2] that: all the derivations, but those of the quantitative part, are done in $\mathcal{L}_{DLAL\star}$. In particular all rules $(\forall e)$ are done on $\mathcal{L}_{DLAL\star}$ formulas. Such a derivation can be converted into a LAL typing derivation for a lambda-term $t$ and it is possible to assume $t$ is in normal form (otherwise we normalize it). Thus, using Proposition 7 we get that all these terms are typable in DLAL. Together with the encoding of polynomials of section 4.3.2 this shows that Ptime Turing machines can be encoded in DLAL. Therefore we have:

**Theorem 20** *If a function $f : \{0,1\}^\star \to \{0,1\}^\star$ is computable in polynomial time, then there exists a lambda-term $t$ and an integer $n$ such that $\vdash_{DLAL} t : W \multimap \S^n W$ and $t$ represents $f$.*

## 5 Discussion on the DLAL type inference problem

As there is a forgetful map from propositional EAL/LAL to simple types (removing modalities and replacing $\multimap$ with $\to$) the problem of type inference for lambda-calculus in these systems can be addressed as a *decoration* problem (in the line of [13]): starting from a simple type for the term, decorate it with modalities in order to obtain a suitable EAL/LAL type. This approach has been explored for EAL ([11]) and LAL ([3, 4]) type inference.

For EAL, types are decorated with sequences in $\{!\}^*$, while for LAL they range over $\{!, \S\}^*$. In both cases the main difficulty is to determine where in the derivation to place the exponentials introduction rules: (! i) for EAL and (! i), ($\S$ i) for LAL. These rules correspond to *boxes* in the proof-nets syntax ([2]).

In [12] an algorithm for EAL type inference was described as follows: first place *abstract boxes* on the simple type derivation, parameterized with integer variables (a box with parameter $n$ corresponds to $n$ ! rules); then express the typing conditions for this *abstract derivation*, which yield linear equations on the parameters. Finding a suitable EAL derivation then amounts to solve these systems of linear equations.

In [4] an analogous method was used for LAL type inference, but as there are here two modalities $\{!, ?\}$ the constraints involved were constraints on words.

The system DLAL corresponds by the $(.)^*$ translation to a fragment of LAL where only $\S^k$ and $!\S^k$ sequences are used (and a certain discipline on ! is enforced). In fact ! and $\S$ are assigned two distinct roles: ! is used to handle potential duplications while $\S$ is used to manage stratification. This suggests carrying out the decoration of the simple type derivation with the following steps:

- step 1: finding non-linear applications; this step deals with placing ! exponentials in the derivation (which is not very different from [13]).

- step 2: completing the type derivation by placing the $\S$ rules, which is then similar to EAL inference.

We leave for future work the proper study of DLAL type inference and of its complexity. A proposal of algorithm following the previous scheme and adapting the EAL procedure of [12] can be found in Appendix G.

## 6  Conclusion and perspectives

We have presented a polymorphic type system for lambda-calculus which guarantees that typed terms can be reduced in a polynomial number of steps, and in polynomial time. This system, DLAL, has been designed as a subsystem of LAL. We have proved that it is complete for the class PTIME by showing how to encode polynomial time Turing machines. Being arguably simpler than Light affine logic, DLAL might help to a better understanding of LAL, in particular of the reduction strategies it induces on lambda-terms. It should also be more amenable to type inference. Other approaches to characterization of complexity classes in lambda-calculus have considered restrictions on type orders (see [15, 21, 24]); it would be interesting to examine the possible relations between this line of work and the present setting based on linear logic. Finally DLAL might provide some new intuitions on the topic of denotational semantics for light logics ([5]).

# APPENDIX

## A  Sequent calculus for LAL

The sequent-calculus presentation of LAL is given on figure 4. It is equivalent to the natural deduction presentation, as a type system:

**Lemma 21** *A judgment $\Gamma \vdash t : A$ is derivable in the LAL sequent calculus iff it is derivable in NLAL.*

## B  Sequent calculus for DLAL

The sequent-calculus presentation of DLAL is given on figure 5.

As usual in a sequent calculus presentation application is handled by the left introduction rule for the arrow connective. Here there are two arrows: $\multimap$ and $\Rightarrow$. Note that in the case of $(\Rightarrow l)$, the argument $u$ is constrained to be typed with a judgment $; z : D \vdash u : A$, so to have at most one variable, which is linear.

Again, it is equivalent to the natural deduction formulation:

**Lemma 22** *A judgment $\Gamma; \Delta \vdash t : A$ is derivable in the DLAL sequent calculus iff it is derivable in NDLAL.*

## C  From derivations in LAL to derivations in DLAL: Proof of Proposition 7

To prove Prop. 7 we first prove the analogous property with sequent calculus typing (Lemma 25) and then use the fact that the sequent calculus and natural deduction presentations are equivalent (Lemmas 21 and 22).

In the rest of this section, unless explicitely stated derivations will be sequent calculus derivations and $\Gamma \vdash_{LAL} t : A$ (resp. $\Gamma; \Delta \vdash_{DLAL} t : A$) will stand for a LAL (resp. DLAL) sequent calculus typing judgment.

**Definition 3** *We say an LAL derivation is* tidy *if it satisfies the following conditions:*

1. *formulas in (Id) rules (axioms) do not start with a ! or §,*

2. *a rule !l introducing a formula !A is followed by a rule in which !A is active ($\forall l$, $\multimap l$, $\multimap r$, Cut, !r, §r) or it is the last rule of the derivation,*

3. *a rule §r is followed by rules §l for all the discharged formulas $[B]_\S$ on the l.h.s. of the sequent, or it is the last rule of the derivation.*

Intuitively: condition 2 says that rules !l are applied *as late as possible* (with top-down orientation); condition 3 that rules §l are applied *as early as possible*.

**Lemma 23 (tidying lemma)** *If $t$ is a lambda-term and $\Gamma \vdash_{LAL} t : A$ is derivable, then this judgement can be obtained with a tidy derivation. If the initial derivation is cut-free, one can give a cut-free tidy derivation.*

*Proof.* If there is in the derivation an $(Id)$ rule (axiom) on a formula of the form $!B$ or $\S B$ then one can $\eta$-expand it, using rules $!l, !r, \S l, \S r$ until getting an $(Id)$ rule which is not of this form.

Then we observe that:

- a !l rule with main formula !A can commute top-down with any rule but one active on !A or rules $!r, \S r$. These commutations do not change the lambda-term associated to the derivation.

- a §l rule acting on $[A]_\S$ can commute top-down with any rule but the one introducing $[A]_\S$, which is necessarily a §r rule. These commutations do not change the lambda-term associated to the derivation.

Applying these commutations we eventually end up with a tidy derivation of the same judgement.

**Lemma 24 (bang lemma)** *If $\mathcal{D}$ is a tidy cut-free LAL derivation of a judgement $[\Gamma]_!, [\Xi]_\S, \Delta \vdash u : !A$ with $\Gamma, \Xi, \Delta, A$ in $\mathcal{L}_{DLAL\star}$, then there exists a derivation $\mathcal{D}'$ of height inferior or equal to that of $\mathcal{D}$ and ending with:*

$$\dfrac{\dfrac{x : B \;\vdash\; u : A}{x : [B]_! \;\vdash\; u : !A} \, !r}{x : [B]_!, \Delta \;\vdash\; u : !A} \, Weak$$

*and we have $\Gamma = B, \Xi = \emptyset$;*

*or the same derivation without $x : B$, in which case we have $\Gamma = \Xi = \emptyset$.*

*Proof.* The r.h.s. !A formula cannot have been introduced by an $(Id)$ rule as the derivation is tidy. Hence it has been introduced by a !r rule. Therefore within $\mathcal{D}$ there is a sub-derivation $\mathcal{D}_1$ ending with a rule:

$$\dfrac{y : C \;\vdash\; t : A}{y : [C]_! \;\vdash\; t : !A} \, !r$$

or the same with no $y : C$ on the l.h.s.

If there is a following rule in $\mathcal{D}$ call it $R$. The rule $R$ can only be a !l or $Weak$ rule. If it is !l it cannot be the last rule, otherwise $\Delta$ would contain a formula $!B$, which does not belong to $\mathcal{L}_{DLAL\star}$. As the derivation is tidy the rule $R$ is followed by a rule active on $!B$: $\forall l, \multimap l, \multimap r, !r, \S r$. The rules $\forall l, \multimap r$ are excluded because they would introduce a

$$\frac{}{x:A \vdash x:A} \; Id \qquad \frac{\Gamma_1 \vdash u:A \quad x:A, \Gamma_2 \vdash t:C}{\Gamma_1, \Gamma_2 \vdash t[u/x]:C} \; Cut$$

$$\frac{\Gamma \vdash t:C}{\Delta, \Gamma \vdash t:C} \; Weak \qquad \frac{x:[A]_!, y:[A]_!, \Gamma \vdash t:C}{z:[A]_!, \Gamma \vdash t[z/x, z/y]:C} \; Cntr$$

$$\frac{\Gamma_1 \vdash u:A_1 \quad x:A_2, \Gamma_2 \vdash t:C}{\Gamma_1, y:A_1 \multimap A_2, \Gamma_2 \vdash t[yu/x]:C} \; \multimap l \qquad \frac{x:A_1, \Gamma \vdash t:A_2}{\Gamma \vdash \lambda x.t:A_1 \multimap A_2} \; \multimap r$$

$$\frac{x:A[B/\alpha], \Gamma \vdash t:C}{x:\forall \alpha.A, \Gamma \vdash t:C} \; \forall l \qquad \frac{\Gamma \vdash t:A}{\Gamma \vdash t:\forall \alpha.A} \; \forall r, \; (\alpha \text{ is not free in } \Gamma)$$

$$\frac{x:[A]_!, \Gamma \vdash t:C}{x:!A, \Gamma \vdash t:C} \; !l \qquad \frac{x:B \vdash t:A}{x:[B]_! \vdash t:!A} \; !r$$

$$\frac{x:[A]_\S, \Gamma \vdash t:C}{x:\S A, \Gamma \vdash t:C} \; \S l \qquad \frac{\Gamma, \Delta \vdash t:A}{[\Gamma]_!, [\Delta]_\S \vdash t:\S A} \; \S r$$

**Figure 4. Sequent-calculus for LAL**

formula not belonging to $\mathcal{L}_{DLAL}$, which is impossible. The rules $\multimap r$, $!r$, $\S r$ are excluded because they would change the r.h.s. formula. Hence the rule $R$ cannot be a $!l$ rule.

Therefore $R$ is a $Weak$ rule. Similarly one can check that if $R$ is not the last rule, then the following rules can only be $Weak$ or $\forall l$, $\multimap l$ acting on weakened formulas. As a consequence we have $y = x$, $C = B$, $t = u$ and one can replace the part of the derivation below $\mathcal{D}_1$ by simply a $Weak$ rule and obtain the same judgement as conclusion. The resulting derivation is $\mathcal{D}'$.

**Lemma 25** *If $t$ is a lambda-term, $\Gamma, \Xi, \Delta, A$ are in $\mathcal{L}_{DLAL}$, and $\mathcal{D}$ is an LAL derivation of the judgement $[\Gamma^*]_!, [\Xi^*]_\S, \Delta^* \vdash t : A^*$ such that:*

- *$\mathcal{D}$ is cut-free,*
- *quantification in $\mathcal{D}$ is only on formulas of $\mathcal{L}_{DLAL\star}$,*

*then $\Gamma; \S\Xi, \Delta \vdash t : A$ is derivable in DLAL.*

*Proof.* To simplify the notations we will omit the symbol $(.)^*$ on formulas when there is no ambiguity.

By lemma 23 one can assume the derivation $\mathcal{D}$ is tidy. Then by the subformula property and the assumption on quantification we get: any formula occurring in $\mathcal{D}$ is in $\mathcal{L}_{DLAL\star}$ or of the form $!A$ with $A$ in $\mathcal{L}_{DLAL\star}$.

We proceed by induction on $\mathcal{D}$, considering its last rule:

- rule $\multimap l$:

  the last rule is of the form:

$$\frac{[\Xi_1]_\S, [\Gamma_1]_!, \Delta_1 \vdash u:B \quad [\Xi_2]_\S, [\Gamma_2]_!, \Delta_2, x:C \vdash t_2:A}{[\Xi]_\S, [\Gamma]_!, \Delta \vdash t_2[y\,u/x]:A} \; \multimap l$$

with $\Gamma = \Gamma_1, \Gamma_2$, $\Delta = \Delta_1, \Delta_2$, $\Xi = \Xi_1, \Xi_2$; call $\mathcal{D}_1$ and $\mathcal{D}_2$ the two immediate subderivations.

As $B \multimap C$ is in $\mathcal{L}_{DLAL\star}$, $C$ is in $\mathcal{L}_{DLAL\star}$. Moreover as $\Xi_2, \Gamma_2, \Delta_2, A \in \mathcal{L}_{DLAL\star}$ one can apply the induction hypothesis to $\mathcal{D}_2$, which gives a DLAL derivation $\mathcal{D}'_2$ of conclusion: $\Gamma_2; \S\Xi_2, \Delta_2, x : C \vdash t_2 : A$.

For $\mathcal{D}_2$ we have two cases:

- **first case**: $B$ is not of the form $!B_1$,

  then $B \in \mathcal{L}_{DLAL}$ and one can apply the i.h. to $\mathcal{D}_1$, getting a DLAL derivation $\mathcal{D}'_1$. We then have a DLAL derivation:

$$\frac{\Gamma_1; \S\Xi_1, \Delta_1 \vdash u:B \quad \Gamma_2; \S\Xi_2, \Delta_2, x:C \vdash t_2:A}{\Gamma; \S\Xi, \Delta, y:B \multimap C \vdash t_2[y\,u/x]:A} \; \multimap l$$

- **second case**: $B = !B_1$, with $B_1 \in \mathcal{L}_{DLAL}$,

  by lemma 24 there exists an LAL derivation $\mathcal{D}_3$ with height inferior to that of $\mathcal{D}_1$ ending with:

$$\frac{\dfrac{z:D_1 \vdash u:B_1}{z:[D_1]_! \vdash u:!B_1} \; !r}{z:[D_1]_!, \Delta_1 \vdash u:!B_1} \; Weak$$

with $\Gamma_1 = D_1, \Xi_1 = \emptyset$,

or

$$\frac{\dfrac{\vdash u:B_1}{\vdash u:!B_1} \; !r}{\Delta_1 \vdash u:!B_1} \; Weak$$

$$\frac{}{;x:A \vdash x:A} \ (Id) \qquad \frac{\Gamma_1;\Delta_1 \vdash u:A \quad \Gamma_2;x:A,\Delta_2 \vdash t:C}{\Gamma_1,\Gamma_2;\Delta_1,\Delta_2 \vdash t[u/x]:C} \ (Cut)$$

$$\frac{\Gamma;\Delta \vdash t:C}{\Sigma,\Gamma;\Pi,\Delta \vdash t:C} \ (Weak) \qquad \frac{x:A,y:A,\Gamma;\Delta \vdash t:C}{z:A,\Gamma;\Delta \vdash t[z/x,z/y]:C} \ (Cntr)$$

$$\frac{\Gamma_1;\Delta_1 \vdash u:A \quad \Gamma_2;x:B,\Delta_2 \vdash t:C}{\Gamma_1,\Gamma_2;y:A \multimap B,\Delta_1,\Delta_2 \vdash t[yu/x]:C} \ (\multimap l) \qquad \frac{\Gamma;x:A,\Delta \vdash t:B}{\Gamma;\Delta \vdash \lambda x.t:A \multimap B} \ (\multimap r)$$

$$\frac{;z:D \vdash u:A \quad \Gamma;x:B,\Delta \vdash t:C}{z:D,\Gamma;y:A \Rightarrow B,\Delta \vdash t[yu/x]:C} \ (\Rightarrow l) \qquad \frac{x:A,\Gamma;\Delta \vdash t:B}{\Gamma;\Delta \vdash \lambda x.t:A \Rightarrow B} \ (\Rightarrow r)$$

$$\frac{;\Gamma,x_1:B_1,\ldots,x_n:B_n \vdash t:A}{\Gamma;x_1:\S B_1,\ldots,x_n:\S B_n \vdash t:\S A} \ (\S)$$

$$\frac{\Gamma;x:A[B/\alpha],\Delta \vdash t:C}{\Gamma;x:\forall\alpha.A,\Delta \vdash t:C} \ (\forall l) \qquad \frac{\Gamma;\Delta \vdash t:A}{\Gamma;\Delta \vdash t:\forall\alpha.A} \ (\forall r), \alpha \text{ is not free in } \Gamma,\Delta$$

**Figure 5. Sequent calculus for DLAL**

with $\Gamma_1 = \emptyset, \Xi_1 = \emptyset$.
Then by i.h. on $\mathcal{D}_3$ we get a DAL derivation $\mathcal{D}'_3$ of either $;z:D_1 \vdash u:B_1$ or $;\vdash u:B_1$. Let us assume for simplicity we are in the first situation (the second one is similar). Then we can take for $\mathcal{D}'$ the following DLAL derivation, starting from subderivations $\mathcal{D}'_3$ and $\mathcal{D}'_2$:

$$\frac{\dfrac{;z:D_1 \vdash u:B_1 \quad \Gamma_2;\S \Xi_2,\Delta_2,x:C \vdash t_2:A}{z:D_1,\Gamma_2;\S \Xi_2,y:!B_1 \multimap C,\Delta_2 \vdash t_2[y\,u/x]:A}\multimap l}{z:D_1,\Gamma_2;\S \Xi_2,y:!B_1 \multimap C,\Delta_1,\Delta_2 \vdash t_2[y\,u/x]:A} Weak$$

- rule $\multimap r$:

  We have $A = B \multimap C$ and the last rule is of the form:

  $$\frac{[\Gamma]_!,\Delta,x:B \vdash t_1:C}{[\Gamma]_!,\Delta \vdash \lambda x.t_1:B \multimap C}\multimap r$$

  with an immediate subderivation that we call $\mathcal{D}_1$.

  We distinguish two cases:

  – **first case**: $B \in \mathcal{L}_{DLAL\star}$,
    then by i.h. on $\mathcal{D}_1$ we get a DLAL derivation $\mathcal{D}'_1$ and complete it in the following way to get $\mathcal{D}'$:

    $$\frac{\Gamma;\Delta,x:B \vdash t_1:C}{\Gamma;\Delta \vdash \lambda x.t_1:B \multimap C}\multimap r$$

  – **second case**: $B = !B_1$ with $B_1 \in \mathcal{L}_{DLAL\star}$,
    as $\mathcal{D}$ is tidy, the $!B_1$ on the l.h.s. has been introduced by a $!l$ rule, which must precede immediately the rule $\multimap r$. Hence $\mathcal{D}$ is of the form:

$$\frac{\dfrac{[\Gamma,B_1]_!,\Delta \vdash t_1:C}{[\Gamma]_!,x:B,\Delta \vdash t_1:C}!l}{[\Gamma]_!,\Delta \vdash \lambda x.t_1:B \multimap C}\multimap r$$

with an immediate subderivation $\mathcal{D}_2$.

By i.h. on $\mathcal{D}_2$ we get a DLAL derivation $\mathcal{D}'_2$, which we complete into a DLAL derivation $\mathcal{D}'$ by:

$$\frac{\Gamma,x:B_1;\Delta \vdash t_1:C}{\Gamma;\Delta \vdash \lambda x.t_1:!B_1 \multimap C}\multimap r$$

- the other inductive cases are straightforward.

*Proof.* [Prop. 7] Assume $t$ is a term in normal form and $[\Gamma']_!,\Delta' \vdash t:A'$ can be derived in NLAL using $(\forall e)$ only with instantiation on $\mathcal{L}_{DLAL\star}$. Then by Lemma 21 there is a LAL sequent calculus derivation $\mathcal{D}$ of $[\Gamma']_!,\Delta' \vdash_{LAL} t:A'$, and quantification in $\mathcal{D}$ is only on $\mathcal{L}_{DLAL\star}$ formulas. As $t$ is in normal form it is easy to see that $\mathcal{D}$ can be taken without cut. Then by Lemma 25 $\Gamma;\Delta \vdash_{DLAL} t:A$ can be derived in DLAL sequent calculus (with $\Gamma^* = \Gamma'$, $\Delta^* = \Delta'$, $A^* = A'$) thus by Lemma 22 in natural deduction DLAL.

## D Proof of subject reduction

### D.1 Proof of lemma 8

*Proof.* (1) By induction on $n$. (2) By induction on $m$. (3) By induction on $m$. When the last rule of the derivation is (§ i):

$$\frac{;\Gamma_2, x:A, \Delta_2' \vdash^{m-1} t:B'}{\Gamma_2; x:[A]_\S, [\Delta_2'] \vdash^m t:\S B'} \;(\S \text{ i})$$

Apply (2) to obtain

$$;\Gamma_1, \Delta_1, \Gamma_2, \Delta_2' \vdash^{n+m-1} t[u/x]:B',$$

then apply ($\S$ i) to obtain

$$\Gamma_1, \Gamma_2; \Delta_1, [\Delta_2'] \vdash^{n+m} t[u/x]:\S B'.$$

(4) By induction on $m$. When the last rule of the derivation is ($\Rightarrow$ e):

$$\frac{\vec{x}:\vec{A}, \Gamma; \Delta \vdash t_1 : D \Rightarrow B \quad ; x_n : A \vdash t_2 : D}{\vec{x}:\vec{A}, x_n : A, \Gamma; \Delta \vdash (t_1\, t_2) : B} \;(\Rightarrow \text{e})$$

where $\vec{x}:\vec{A} \equiv x_1 : A_1, \ldots, x_{n-1} : A_{n-1}$. By the induction hypothesis, we have

$$z:C, \Gamma; \Delta \vdash t_1[u/\vec{x}] : D \Rightarrow B,$$

while by (2), we also have

$$; z:C \vdash t_2[u/x_n] : D.$$

¿From these two, we immediately obtain the desired result:

$$z:C, \Gamma; \Delta \vdash (t_1[u/\vec{x}]\, t_2[u/x_n]) : D \Rightarrow B.$$

## D.2 Proof of Lemma 9

*Proof.* When the first or the second condition is violated, apply the following rewriting rules:

$$\frac{\begin{array}{c}\vdots \mathcal{D} \\ \Gamma; \Delta \vdash^n t:A \\ \hline \Gamma; \Delta \vdash^{n+1} t:\forall \alpha.A \end{array} (\forall \text{i})}{\Gamma; \Delta \vdash^{n+2} t:A[B/\alpha]} (\forall \text{e}) \;\Longrightarrow\; \begin{array}{c}\vdots \mathcal{D}' \\ \Gamma; \Delta \vdash^n t:A[B/\alpha]\end{array}$$

$$\frac{\begin{array}{cc}\vdots \mathcal{D}_1 & \\ ;\Gamma_1, \Delta_1 \vdash^n u:A & \vdots \mathcal{D}_2 \\ \hline \Gamma_1; [\Delta_1]_\S \vdash^{n+1} u:\S A \end{array} (\S \text{i}) \quad \Gamma_2; x:[A]_\S, \Delta_2 \vdash^m t:B}{\Gamma_1, \Gamma_2; [\Delta_1]_\S, \Delta_2 \vdash^{n+m+2} t[u/x]:B} (\S \text{e})$$

$$\Longrightarrow \begin{array}{c}\vdots \mathcal{D}'' \\ \Gamma_1, \Gamma_2; [\Delta_1]_\S, \Delta_2 \vdash^{n+m} t[u/x]:B \end{array}$$

where $\mathcal{D}'$ and $\mathcal{D}''$ are derivations obtained by Substitution Lemma (1) and (3) respectively. The size of the derivation strictly decreases. When the third condition is violated, permute the two rules at issue: for instance, when the conclusion of a ($\S$ e) rule is the major premise of another ($\S$ e) rule, apply the rewriting rule in Figure 6. It is not hard to see that, given a derivation, the process of applying the above rewriting rules terminates eventually, resulting in a $\forall\S$-normal derivation.

## E Simulation lemma and polynomial time strong normalization

In this section, we will give a simulation of DLAL typable lambda terms by terms of $\lambda$LA. More specifically, we show that every DLAL typable lambda term $t$ translates to a term $\tilde{t}$ of $\lambda$LA (depending on the typing derivation for $t$), and that any beta reduction sequence from $t$ can be simulated by a *longer* $\lambda$LA reduction sequence from $\tilde{t}$. The polynomial time strong normalization theorem for DLAL directly follows from this fact.

Let us first recall light affine lambda calculus $\lambda$LA from [25].

**Definition 4** *The set of (pseudo) terms of $\lambda$LA is defined by the following grammar:*

$$t, u ::= x \mid \lambda x.t \mid tu \mid !t \mid \text{let } u \text{ be } !x \text{ in } t \mid \S t \mid \text{let } u \text{ be } \S x \text{ in } t.$$

*A term of the form $(\lambda x.\text{let } x \text{ be } !y \text{ in } t[y/x])$, where $y$ is fresh, is abbreviated by $\lambda^! x.t$.*

The *depth* of $t$ is the maximal number of occurrences of $!u$ and $\S u$ in a branch of the term tree for $t$.

DLAL can be considered as a type system for $\lambda$LA. We write $\Gamma; \Delta \vdash^{\lambda\text{LA}}_{DLAL} t : A$ if $t$ is a term of $\lambda$LA and $\Gamma; \Delta \vdash t : A$ is derivable by the type assignment rules in Figure 7. The *depth* of a DLAL derivation $\mathcal{D}$ is the maximal number of premises of ($\S$ i) and r.h.s. premises of ($\Rightarrow$ e) in a branch of $\mathcal{D}$.

The reduction rules of $\lambda$LA are given on Figure 8.

A term $t$ is $(\S, !, com)$-*normal* if neither of the reduction rules $(\S)$, $(!)$, $(com1)$ and $(com2)$ applies to $t$. We write $t \xrightarrow{(\beta^*)} u$ when $t$ reduces to $u$ by $(\beta)$ followed by several applications of $(\S)$, $(!)$, $(com1)$ and $(com2)$. Given an $\lambda$LA-term $t$, its *erasure* $t^-$ is defined by:

$$\begin{array}{rclrcl} x^- & \equiv & x & (tu)^- & \equiv & t^- u^- \\ (\lambda x.t)^- & \equiv & \lambda x.(t^-) & (\dagger t)^- & \equiv & t^- \\ (\text{let } u \text{ be } \dagger x \text{ in } t)^- & \equiv & t^-[u^-/x] \end{array}$$

The following is the main result of [25]:

**Theorem 26 (Polytime strong normalization for $\lambda$LA)**
*Any typable $\lambda$LA-term $t$ of depth $d$ reduces to the normal form in $O(|t|^{2^{d+1}})$ reduction steps, and in time $O(|t|^{2^{d+2}})$ on a Turing machine. This result holds independently of which reduction strategy we take.*

**Lemma 27 (DLAL and $\lambda$LA)**

$$
\begin{array}{c}
\vdots\mathcal{D}_1 \qquad \vdots\mathcal{D}_2 \\
\dfrac{\Gamma_1;\Delta_1 \vdash v:\S A \quad \Gamma_2;x:[A]_\S,\Delta_2 \vdash u:\S B}{\Gamma_1,\Gamma_2;\Delta_1,\Delta_2 \vdash u[v/x]:\S B} \ (\S\ e) \qquad \dfrac{\vdots\mathcal{D}_3}{\Gamma_3;y:[B]_\S,\Delta_3 \vdash t:C} \\
\dfrac{}{\Gamma_1,\Gamma_2,\Gamma_3;\Delta_1,\Delta_2,\Delta_3 \vdash t[u[v/x]/y]:C} \ (\S\ e)
\end{array}
$$

$$\Downarrow$$

$$
\begin{array}{c}
\vdots\mathcal{D}_2 \qquad \vdots\mathcal{D}_3 \\
\vdots\mathcal{D}_1 \qquad \dfrac{\Gamma_2;x:[A]_\S,\Delta_2 \vdash u:\S B \quad \Gamma_3;y:[B]_\S,\Delta_3 \vdash t:C}{\Gamma_2,\Gamma_3;x:[A]_\S,\Delta_2,\Delta_3 \vdash t[u/y]:C} \ (\S\ e) \\
\dfrac{\Gamma_1;\Delta_1 \vdash v:\S A \qquad\qquad\qquad\qquad\qquad\qquad\qquad}{\Gamma_1,\Gamma_2,\Gamma_3;\Delta_1,\Delta_2,\Delta_3 \vdash t[u/y][v/x]:C} \ (\S\ e)
\end{array}
$$

**Figure 6. Rewriting rule**

---

$$\dfrac{}{;x:A \vdash x:A} \ \text{(variable)}$$

$$\dfrac{\Gamma_1;\Delta_1,x:A \vdash t:B}{\Gamma_1;\Delta_1 \vdash \lambda x.t:A \multimap B} \ (\multimap\ i) \qquad \dfrac{\Gamma_1;\Delta_1 \vdash t:A \multimap B \quad \Gamma_2;\Delta_2 \vdash u:A}{\Gamma_1,\Gamma_2;\Delta_1,\Delta_2 \vdash (t\ u):B} \ (\multimap\ e)$$

$$\dfrac{\Gamma_1,x:A;\Delta_1 \vdash t:B}{\Gamma_1;\Delta_1 \vdash \lambda^! x.t:A \Rightarrow B} \ (\Rightarrow\ i) \qquad \dfrac{\Gamma_1;\Delta_1 \vdash t:A \Rightarrow B \quad ;z:C \vdash u:A}{\Gamma_1,z:C;\Delta_1 \vdash (t\ !u):B} \ (\Rightarrow\ e)$$

$$\dfrac{\Gamma_1;\Delta_1 \vdash t:A}{\Gamma_1,\Gamma_2;\Delta_1,\Delta_2 \vdash t:A} \ \text{(Weak)} \qquad \dfrac{x_1:A,x_2:A,\Gamma_1;\Delta_1 \vdash t:B}{x:A,\Gamma_1;\Delta_1 \vdash t[x/x_1,x/x_2]:B} \ \text{(Cntr)}$$

$$\dfrac{;\Delta_1,\Delta_2 \vdash t:A}{\Delta_1;[\Delta_2]_\S \vdash \S t:\S A} \ (\S\ i) \qquad \dfrac{\Gamma_1;\Delta_1 \vdash u:\S A \quad \Gamma_2;x:[A]_\S,\Delta_2 \vdash t:B}{\Gamma_1,\Gamma_2;\Delta_1,\Delta_2 \vdash \mathsf{let}\ u\ \mathsf{be}\ \S x\ \mathsf{in}\ t:B} \ (\S\ e)$$

$$\dfrac{\Gamma_1;\Delta_1 \vdash t:A}{\Gamma_1;\Delta_1 \vdash t:\forall \alpha.A} \ (\forall\ i)\ (*) \qquad \dfrac{\Gamma_1;\Delta_1 \vdash t:\forall \alpha.A}{\Gamma_1;\Delta_1 \vdash t:A[B/\alpha]} \ (\forall\ e)$$

**Figure 7. DLAL as a type system for $\lambda$LA**

---

(1) If $\Gamma;\Delta \vdash^{\lambda\text{LA}}_{DLAL} t : A$, then $\Gamma;\Delta \vdash_{DLAL} t^- : A$ and $|t^-| \leq |t|$.

(2) If $\Gamma;\Delta \vdash_{DLAL} t : A$, then there is a $\lambda$LA-term $\tilde{t}$ such that $\Gamma;\Delta \vdash^{\lambda\text{LA}}_{DLAL} \tilde{t} : A$ is derivable, $(\tilde{t})^- \equiv t$, and the size and depth of $\tilde{t}$ are bounded by those of the derivation of $\Gamma;\Delta \vdash_{DLAL} t : A$.

*Proof.* By induction on the derivation.

**Lemma 28** *Let $t$ be a term of $\lambda$LA which is neither a variable $x$, application $(u\ v)$ nor let $u$ be $\S x$ in $v$.*

(1) *If $\Gamma;\Delta \vdash^{\lambda\text{LA}}_{DLAL} t : \forall \alpha_1 \cdots \forall \alpha_n.A \multimap B\ (n \geq 0)$, then $t$ is of the form $\lambda x.u$.*

(2) *If $\Gamma;\Delta \vdash^{\lambda\text{LA}}_{DLAL} t : \forall \alpha_1 \cdots \forall \alpha_n.A \Rightarrow B\ (n \geq 0)$, then $t$ is of the form $\lambda^! x.u$.*

(3) *If $\Gamma;\Delta \vdash^{\lambda\text{LA}}_{DLAL} t : \forall \alpha_1 \cdots \forall \alpha_n.\S A\ (n \geq 0)$ is derivable, then $t$ is of the form $\S u$.*

*Proof.* By induction on the derivation.

**Lemma 29**

(1) *If $\Gamma;\Delta \vdash^{\lambda\text{LA}}_{DLAL} (t\ u) : A$ and $(t\ u)$ is $(\S,!,com)$-normal, then $t$ is either $x$, $(v_1\ v_2)$ or $\lambda x.v$.*

(2) *If $\Gamma;\Delta \vdash^{\lambda\text{LA}}_{DLAL}$ let $t$ be $\S x$ in $u$ : $A$ and let $t$ be $\S x$ in $u$ is $(\S,!,com)$-normal, then $u$ is either $x$ or $(v_1\ v_2)$.*

*Proof.* (1) Assume that $t$ is neither $x$ nor $(u_1\ u_2)$. The proof is carried out by induction on the derivation. If the last inference rule is $(\multimap r)$ of the form:

$$\dfrac{\Gamma_1;\Delta_1 \vdash t:A \multimap B \quad \Gamma_2;\Delta_2 \vdash u:A}{\Gamma_1,\Gamma_2;\Delta_1,\Delta_2 \vdash (t\ u):B} \ (\multimap\ e)$$

then $t$ cannot be of the form let $v_1$ be $\S x$ in $v_2$ since $(t\ u)$ is $(com)$-normal. Hence by Lemma 28 (1), $t$ is an abstraction. The other cases are similar.

$$
\begin{array}{lrcl}
(\beta) & (\lambda x.t)u & \longrightarrow & t[u/x] \\
(\S) & \text{let } \S u \text{ be } \S x \text{ in } t & \longrightarrow & t[u/x] \\
(!) & \text{let } !u \text{ be } !x \text{ in } t & \longrightarrow & t[u/x] \\
(com1) & (\text{let } u \text{ be } \dagger x \text{ in } t)v & \longrightarrow & \text{let } u \text{ be } \dagger x \text{ in } (tv) \\
(com2) & \text{let } (\text{let } u \text{ be } \dagger x \text{ in } t) \text{ be } \dagger y \text{ in } v & \longrightarrow & \text{let } u \text{ be } \dagger x \text{ in } (\text{let } t \text{ be } \dagger y \text{ in } v)
\end{array}
$$

**Figure 8. Reduction rules of $\lambda$LA**

(2) Assume that $t$ is neither $x$ nor $(u_1\ u_2)$. The proof is again by induction on the derivation. If the last rule is

$$\frac{\Gamma_1; \Delta_1 \vdash t : \S A \quad \Gamma_2; x : [A]_\S, \Delta_2 \vdash u : B}{\Gamma_1, \Gamma_2; \Delta_1, \Delta_2 \vdash \text{let } t \text{ be } \S x \text{ in } u : B} \ (\S\, e)$$

then $t$ cannot be of the form let $v_1$ be $\S x$ in $v_2$ since $t$ is $(com)$-normal. Hence by Lemma 28 (3), $t$ must be of the form $\S v$, but that is impossible since let $t$ be $\S x$ in $u$ is $(\S)$-normal. The other cases are immediate.

**Lemma 30 (Simulation)** *Let $t$ be a term of $\lambda$LA which is a subterm of a typable term and $(\S, !, com)$-normal. If $t^-$ reduces to $u$ by $(\beta)$ reduction, then there is a $(\S, !, com)$-normal term $\tilde{u}$ of $\lambda$LA such that $t \xrightarrow{(\beta^*)}_* \tilde{u}$ and $(\tilde{u})^- \equiv u$:*

$$
\begin{array}{ccc}
t^- & \xrightarrow{(\beta)} & u \\
\uparrow{\scriptstyle -} & & \vdots{\scriptstyle -} \\
t & \xdashrightarrow{(\beta^*)} & \tilde{u}
\end{array}
$$

*Proof.* By induction on $t$.
(Case 1) $t$ is a variable. Trivial.
(Case 2) $t$ is of the form $\lambda x.v$. By the induction hypothesis.
(Case 3) $t$ is of the form $(u_1\ u_2)$. In this case, $t^-$ is $(u_1^-\ u_2^-)$. When the redex is inside $u_1^-$ or $u_2^-$, the induction hypothesis applies. When the redex is $t^-$ itself, then $u_1^-$ must be of the form $\lambda x.v$. By the definition of erasure, $u_1$ cannot be a variable nor an application. Therefore, by Lemma 29 (1), $u_1$ must be of the form $\lambda x.\tilde{v}$ with $(\tilde{v})^- \equiv v$. We therefore have

$$
\begin{array}{ccc}
(\lambda x.v)u_2^- & \xrightarrow{(\beta)} & v[u_2^-/x] \\
\uparrow{\scriptstyle -} & & \vdots{\scriptstyle -} \\
(\lambda x.\tilde{v})u_2 & \xdashrightarrow{(\beta^*)} & \tilde{v}[u_2/x]
\end{array}
$$

as required.
(Case 4) $t$ is of the form $!v$. By the induction hypothesis.
(Case 5) $t$ is of the form let $u_1$ be $!x$ in $u_2$. Since $t$ is a subterm of a term typable in DLAL, $u_1$ must be a variable $y$. Therefore, $t^-$ is of the form $u_2^-[y/x]$. It is then not hard to see that if $t^-$ reduces to $u$, there is some $u'$ such that $u_2^- \longrightarrow u'$ and $u \equiv u'[y/x]$. By the induction hypothesis, there is $\tilde{u}$ such that

$$
\begin{array}{ccc}
u_2^- & \xrightarrow{(\beta)} & u' \\
\uparrow{\scriptstyle -} & & \vdots{\scriptstyle -} \\
u_2 & \xdashrightarrow{(\beta^*)} & \tilde{u}'
\end{array}
$$

We therefore have

$$
\begin{array}{ccc}
u_2^-[y/x] & \xrightarrow{(\beta)} & u'[y/x] \\
\uparrow{\scriptstyle -} & & \vdots{\scriptstyle -} \\
\text{let } y \text{ be } !x \text{ in } u_2 & \xdashrightarrow{(\beta^*)} & \text{let } y \text{ be } !x \text{ in } \tilde{u}'
\end{array}
$$

as required.
(Case 6) $t$ is of the form $\S v$. By the induction hypothesis.
(Case 7) $t$ is of the form let $u_1$ be $\S x$ in $u_2$. In this case, $t^-$ is $u_2^-[u_1^-/x]$. By Lemma 29 (2), $u_1$ is either a variable or an application, and so is $u_1^-$. Therefore, the redex in $t$ is either inside $u_1^-$ or results from a redex in $u_2^-$ by substituting $u_1^-$ for $x$. In the latter case, the proof is similar to that of (Case 5). In the former case, let $u_1^- \longrightarrow u$. Then by the induction hypothesis, there is some $\tilde{u}$ such that

$$
\begin{array}{ccc}
u_1^- & \xrightarrow{(\beta)} & u \\
\uparrow{\scriptstyle -} & & \vdots{\scriptstyle -} \\
u_1 & \xdashrightarrow{(\beta^*)} & \tilde{u}
\end{array}
$$

Therefore, we have

$$
\begin{array}{ccc}
u_2^-[u_1^-/x] & \xrightarrow{(\beta)} & u_2^-[u/x] \\
\uparrow{\scriptstyle -} & & \vdots{\scriptstyle -} \\
\text{let } u_1 \text{ be } \S x \text{ in } u_2 & \xdashrightarrow{(\beta^*)} & \text{let } \tilde{u} \text{ be } \S x \text{ in } u_2
\end{array}
$$

as required.

**Theorem 31 (Polynomial time strong normalization)**
*Let $t$ be a $\lambda$-term which has a typing derivation $\mathcal{D}$ in DLAL.*

*Suppose that $\mathcal{D}$ be of size $n$ and of depth $d$. Then $t$ reduces to the normal form $u$ in $O(n^{2^{d+1}})$ reduction steps and in time $O(n^{2^{d+2}})$ on a Turing machine. This result holds independently of which reduction strategy we take.*

*Proof.* By Lemma 27 (2), there is a term $\tilde{t}$ of $\lambda$LA such that $(\tilde{t})^- \equiv t$ and $|\tilde{t}|$ is bounded by the size of $\mathcal{D}$. Hence by Lemma 30, we have:

$$\begin{array}{ccc} t & \xrightarrow{(\beta)} \cdots\cdots \xrightarrow{(\beta)} & u \\ \uparrow^{-} & & \vdots^{-} \\ \tilde{t} & \xrightarrow{(\beta^*)} \cdots\cdots \xrightarrow{(\beta^*)} & \tilde{u} \end{array}$$

Since the length of the reduction sequence from $\tilde{t}$ to $\tilde{u}$ is bounded by $O(|\tilde{t}|^{2^{d+1}}) \leq O(|\mathcal{D}|^{2^{d+1}})$, so is the one from $t$ to $u$.

## F  Normalization

### F.1  Proof of Lemma 17

Let us temporarily use an explicit substitution notation $t\{u/x\}$, and call a stratified term with explicit substitution notations an *x-term*. The variable $x$ is *bound* in $t\{u/x\}$, and the standard variable convention is adopted for explicit substitution notations as well. There is an obvious map $(.)^-$ from the x-terms to the original stratified term, given by $(t\{u/x\})^- = t^-[u^-/x]$. In the following, $t\theta$ stand for an x-term of the form $t\{u_1/x_1\}\cdots\{u_n/x_n\}$.

We prove the following by induction on the number of reduction steps: whenever $t \xrightarrow{d}{}^* u$, there is an x-term $\tilde{u}$ such that

(1) $(\tilde{u})^- = u$,

(2) $|\tilde{u}| \leq |t|$, and

(3) if either $(\lambda x^{d!}.u_1)\theta u_2$ or $u_1\{u_2/x\}$ occurs in $\tilde{u}$, then $u_2$ contains neither a redex at depth $d$ nor an explicit substitution; furthermore, $u_2$ may have at most one free variable, and in case it has, that variable is either free in $\tilde{u}$ or is bound by an abstraction of the form $\lambda^{d!}y.v$.

In the base case, we take $\tilde{u} \equiv t$. The third property is easily checked by induction on the size of a $\forall\S$-normal typing derivation for $t$. In other cases, we simulate beta reduction by the following reduction rules on x-terms:

$$\begin{array}{rcl} (\lambda x^{d!}.t)\theta u & \longrightarrow & (t\theta)\{u/x\} \\ (\lambda x^d.t)\theta u & \longrightarrow & t[u/x]\theta. \end{array}$$

It is easily checked that these reduction rules preserve the above properties.

Let us denote by $no(u)$ the number of free variable occurrences in $u$. We now prove the following by induction on the structure of $\tilde{u}$: when $|\tilde{u}| \geq 2$,

(4) $no(\tilde{u}^-) \leq |\tilde{u}|$, and

(5) $|\tilde{u}^-| \leq |\tilde{u}| \cdot (|\tilde{u}| - 1)$.

Suppose $\tilde{u} \equiv u_1\{u_2/x\}$. Then (4) holds since

$$\begin{array}{rcl} no((u_1\{u_2/x\})^-) & \leq & no(u_1^-) - no(x, u_1^-) + \\ & & no(x, u_1^-) \cdot no(u_2^-) \\ & \leq & |u_1| - no(x, u_1^-) + no(x, u_1^-) \cdot 1 \\ & \leq & |u_1| \leq |\tilde{u}|, \end{array}$$

by the induction hypothesis and (3) above (since $u_2^- \equiv u_2$ and $no(u_2) \leq 1$). As for (5), if $u_1$ is a variable, then $|(u_1\{u_2/x\})^-| \leq |u_2|$, hence the claim holds trivially. Otherwise, $|u_1| \geq 2$ and we can use the induction hypothesis on $u_1$ (in addition to (4)). Thus,

$$\begin{array}{rcl} |(u_1\{u_2/x\})^-| & \leq & |u_1^-| + no(x, u_1^-) \cdot |u_2^-| \\ & \leq & |u_1| \cdot (|u_1| - 1) + |u_1| \cdot |u_2| \\ & \leq & |u_1| \cdot (|u_1| - 1 + |u_2|) \\ & \leq & |\tilde{u}|(|\tilde{u}| - 1). \end{array}$$

Putting (1), (2) and (5) together, we have $|u| = |(\tilde{u})^-| \leq |\tilde{u}|(|\tilde{u}| - 1) \leq |t|(|t| - 1)$ whenever $|u| \geq 2$.

## G  Type inference

One advantage of DLAL over LAL is that it assigns two distinct roles to ! and $\S$: the modality ! is used to handle potential duplications while $\S$ is used to manage stratification. This separation shows up in particular with type-inference, where in the case of DLAL we can take care of the two modalities one at a time (contrarily to what happens with LAL).

We give here a type-inference algorithm for propositional DLAL, which starting from a lambda-term $t$ and its principal simple type $B$ finds all possible decorations of $B$ (if any) into a valid DLAL type for $t$. It will use a type-inference procedure for Elementary affine logic (EAL). Type-inference algorithms for EAL have been given in [11, 12]. Here we will use the algorithm of [12].

Given $t$ and its principal simple type $B$, with environment $\Gamma$ for the free variables, we will try to decorate the simple type derivation $\mathcal{D}$ of $\Gamma \vdash t : B$ into a LAL derivation corresponding to a DLAL derivation (by the $(.)^*$ translation). For that we proceed in two stages:

- **stage 1**: non-linear arguments stage;

    in this stage we place the ! rules in the derivation. This corresponds to working out which arguments are linear

and which arguments are non-linear. It is close to the problem of linear decoration of intuitionistic derivations studied in [13].

- **stage 2**: stratification stage;

  in this stage we complete the type derivation by placing § rules; for that we use the EAL type-inference procedure.

All solutions found by the procedure will give valid DLAL type derivations for $t$. Conversely if $t$ can be typed in DLAL with a judgement $\Delta \vdash_{DLAL} t : C$ which is a decoration of $\Gamma \vdash t : B$, then the procedure will provide a derivation of $\Delta \vdash_{DLAL} t : C$.

We adopt the following conventions for the simple type derivation $\mathcal{D}$ of $\Gamma \vdash t : B$: environments are handled as multisets; application requires both terms to have environments with disjoint sets of variables; contraction and weakening are handled with explicit rules (with a substitution by a fresh variable for contraction (Cntr)) and are performed only just before doing an abstraction on the variable.

**Stage 1:** non-linear arguments stage.

We need to determine which applications of the term should correspond to ($\multimap$ e) or to ($\Rightarrow$ e) rules, which is tied to the issue of working out which abstractions correspond to ($\multimap$ i) or to ($\Rightarrow$ i) rules.

For that we will associate a boolean parameter to each application and abstraction rule of the derivation $\mathcal{D}$, decorate accordingly the types with these parameters and express the validity of this *abstract* derivation by some constraints which should be satisfied.

We consider a set of parameters $a, b \ldots$ meant to range over $\{0, 1\}$. The value $a = 1$ corresponds in a type to a ! modality, while $a = 0$ corresponds to absence of ! modality.

The constraints are of the form: $d_1 = d_2$, where $d_i$ is either a disjunction of parameters $a_1 \vee \cdots \vee a_n$ or a constant 0 or 1. For convenience we will denote here $a_1 \ldots a_n$ for $a_1 \vee \cdots \vee a_n$ and use notation $u, v \ldots$ for such disjunctions, with $n \geq 0$..

*Abstract types* are defined by the two grammars:

$$B ::= \alpha \mid A \to B$$
$$A ::= a_1 \ldots a_n B$$

where $n \geq 0$ and $a_1, \ldots, a_n$ are any parameters. The $B$s are called *basic abstract types*.

An *instantiation* $\phi$ is a map from parameters to $\{0, 1\}$. We write $\phi_1 \leq \phi_2$ if for any parameter $a$ we have $\phi_1(a) \leq \phi_2(a)$.

Let $U(A_1, A_2)$ be the set of constraints on parameters obtained for unifying two abstract types $A_1$ and $A_2$, defined on Figure 9. If $A_1$ and $A_2$ are abstract types with same underlying simple type, then $m(A_1, A_2)$ is defined inductively by: $m(A_1, A_2) = u_1 u_2 \alpha$ if $A_i = u_i \alpha$ for $i = 1, 2$;

$m(A_1, A_2) = u_1 u_2 (m(A'_1, A'_2) \to m(B_1, B_2))$ if $A_i = u_i(A'_i \to B_i)$ for $i = 1, 2$.

We handle *abstract judgements* of the following form: $\Gamma \vdash t : B$ where $B$ is a basic abstract type, $\Gamma$ is a environment assigning abstract types to variables.

If $\Gamma$ is an environment, the notation $a\Gamma$ will stand for the environment given by: $a\Gamma(x)$ is defined iff $\Gamma(x) = A$ is defined, and then $a\Gamma(x) = aA$.

A *maximal decoration* $\overline{A}$ of a simple type $A$ is a basic abstract type defined by induction on $A$ in the following way: if $A = \alpha$ atomic then $\overline{A} = \alpha$, if $A = A_1 \to A_2$ then $\overline{A} = (a\overline{A_1}) \to \overline{A_2}$ where the $\overline{A_i}$ are maximal decorations with disjoint parameters and $a$ is a fresh parameter.

Given a simple type derivation $\mathcal{D}$ we will define inductively a derivation of abstract judgments $\overline{\mathcal{D}}$ and a set of constraints $\mathcal{C}(\mathcal{D})$. Basically the idea is to add a parameter to each argument of application and to each abstraction in order to determine which abstractions should be non-linear.

Given $\mathcal{D}$, $\overline{\mathcal{D}}$ and $\mathcal{C}(\mathcal{D})$ are defined by:

- if $\mathcal{D}$ is just an axiom rule $x : A \vdash x : A$ then $\overline{\mathcal{D}}$ is obtained by replacing $A$ by a maximal decoration $\overline{A}$ and $\mathcal{C}(\mathcal{D}) = $ true, the empty set of constraints.

- if $\mathcal{D}$ is obtained by an application rule on $\mathcal{D}_1$ and $\mathcal{D}_2$, then $\overline{\mathcal{D}}$ is defined from $\overline{\mathcal{D}}_1$ and $\overline{\mathcal{D}}_2$ (taken with disjoint parameters) using a fresh parameter $a$ with the (app $a$) rule of Figure 10. We set $\mathcal{C}(\mathcal{D}) = \mathcal{C}(\mathcal{D}_1) \cup \mathcal{C}(\mathcal{D}_2) \cup U(A_1, aA_2)$.

- if $\mathcal{D}$ is obtained by an abstraction rule on $\mathcal{D}_1$ define similarly $\overline{\mathcal{D}}$ from $\overline{\mathcal{D}}_1$ using the (abstr $a$) rule of Figure 10. We set

$$\mathcal{C}(D) = \begin{cases} \mathcal{C}(D_1) & \text{if } no(x,t) \leq 1, \\ \mathcal{C}(D_1) \cup \{a = 1\} & \text{if } no(x,t) \geq 2. \end{cases}$$

- if $\mathcal{D}$ is obtained by a contraction rule on $\mathcal{D}_1$ define $\overline{\mathcal{D}}$ from $\overline{\mathcal{D}}_1$ using the (Cntr) rule of Figure 10.

- if $\mathcal{D}$ is obtained from $\mathcal{D}_1$ by a weakening rule, then $\overline{\mathcal{D}}$ has as last rule a weakening on a maximal decoration formula.

We now come back to the simple type derivation $\mathcal{D}$ of $t$ and consider the associated abstract derivation $\overline{\mathcal{D}}$ and constraints $\mathcal{C}(\mathcal{D})$, that we will denote as $\mathcal{C}$. Note that $\mathcal{C}$ has at least one solution, as the constant instantiation $\phi \equiv 1$ is a solution.

From a solution $\phi$ and the abstract derivation $\overline{\mathcal{D}}$ one defines a !-*derivation* $\tilde{\mathcal{D}}$: $\tilde{\mathcal{D}}$ is the derivation $\mathcal{D}$ where application rules corresponding to (app $a$) with $\phi(a) = 1$ are annotated as ($\Rightarrow e$) (note that the types themselves are unchanged). In $\tilde{\mathcal{D}}$ we say (thinking about LAL proof-nets) that the r.h.s. subderivation above an ($\Rightarrow e$) rule is in a !-*box*.

$$U(a_1 \ldots a_n(A_1 \to B_1), b_1 \ldots b_m(A_2 \to B_2)) = \{a_1 \vee \cdots \vee a_n = b_1 \vee \cdots \vee b_m\} \cup U(A_1, A_2) \cup U(B_1, B_2)$$
$$U(a_1 \ldots a_n \alpha, b_1 \ldots b_m \alpha) = \{a_1 \vee \cdots \vee a_n = b_1 \vee \cdots \vee b_m\}$$
$$U(A, A') = \text{false} \quad \text{in the other cases.}$$

**Figure 9. Unification of abstract types**

$$\frac{\Gamma_1, x : A \vdash t : B}{\Gamma_1 \vdash \lambda x.t : (aA) \to B} \text{ (abstr } a\text{)} \qquad \frac{\Gamma_1 \vdash t_1 : A_1 \to B_1 \quad \Gamma_2 \vdash t_2 : A_2}{\Gamma_1, a\Gamma_2 \vdash (t_1\, t_2) : B_1} \text{ (app } a\text{)}$$

$$\frac{x_1 : A_1, x_2 : A_2, \Gamma \vdash t : B}{x : m(A_1, A_2), \Gamma \vdash t[x/x_1, x/x_2] : B} \text{ (Cntr)}$$

**Figure 10. Rules for abstract derivations**

We will try to decorate a !-derivation $\tilde{\mathcal{D}}$ (coming from a solution $\phi$) into a DLAL derivation if the following necessary conditions are satisfied:

(i) in $\tilde{\mathcal{D}}$ any r.h.s. premise of a $(\Rightarrow e)$ rule has an environment with at most one variable,

(ii) a variable belongs to at most one environment of r.h.s. premise of $(\Rightarrow e)$.

These conditions are necessary for being able to decorate the derivation into a DLAL derivation; in particular (ii) is needed to ensure that the variable in a r.h.s. environment of $(\Rightarrow e)$ is linear, in the DLAL derivation.

If no solution $\phi$ gives a !-derivation satisfying (i) and (ii) then the initial simple type cannot be decorated into a DLAL type. If some solutions satisfy (i) and (ii) then we try to decorate the corresponding derivations $\tilde{\mathcal{D}}$ into DLAL derivations with stage 2 of the procedure.

**Stage 2:** stratification stage.

Assume $\tilde{\mathcal{D}}$ is a !-derivation obtained by stage 1 and satisfying (i) and (ii).

Let us briefly recall the EAL type inference procedure of [12]. First we recall the notion of *type schemes*. We consider parameters $n, m, n_1, \ldots$ ranging over $\mathbb{N}^\star$. Type schemes are defined by the grammar:

$$\sigma, \sigma' ::= \alpha \,|\, \sigma \multimap \sigma' \,|\, !^{n_1 + \cdots + n_k}\sigma$$

where $k$ can take any positive value and $n_1, \ldots n_k$ are parameters.

The EAL type inference procedure starts from a lambda-term $t$ and proceeds in 3 steps:

- from the term $t$ a set $C(t)$ of *canonical simple forms* of $t$ is computed. A *canonical simple form* of $t$ is a kind of EAL meta-derivation corresponding to $t$. The set $C(t)$ is finite.

- an algorithm $PT(.)$ computes, given a canonical simple form $Q$, a triple $PT(Q) = <\theta, \sigma, \mathcal{C}>$ where: $\theta$ is an assignment of types to variables, $\sigma$ is a type scheme for $Q$ and $\mathcal{C}$ is a set of linear equations on parameters (constraints).

- for any canonical simple form $Q$ of $C(t)$, if $PT(Q) = <\theta, \sigma, \mathcal{C}>$ and $\mathcal{C}$ has a solution $X$, then from $Q, \theta, \sigma$ an EAL type derivation for $t$ can be constructed.

It was shown that this algorithm is correct and complete for EAL (with respect to the EAL typing system without sharing: contraction is allowed only on variables).

In stage 2 of our procedure we proceed in the following way:

- a) first we apply the previous method to $t$ to get its set $C(t)$ of canonical simple forms;

- b) among $C(t)$ we then determine a subset $\tilde{C}(t)$ of canonical simple forms compatible with $\tilde{\mathcal{D}}$;

- c) we apply function $PT(.)$ to the elements of $\tilde{C}(t)$. If for $Q$ in $\tilde{C}(t)$ we have $PT(Q) = <\theta, \sigma, \mathcal{C}>$ and there is a solution $X$ to $\mathcal{C}$, then from $Q, \tilde{\mathcal{D}}, \theta, \sigma$ a DAL type derivation for $t$ can be constructed.

Let us make explicit these steps. To a canonical simple form $Q$ one can associate a syntactic tree with boxes $\mathcal{T}$ (the boxes correspond to the $\nabla$ constructors of the canonical simple form). When naming boxes we will use $\mathcal{B}, \mathcal{B}_1 \ldots$. If we forget about the boxes the syntactic tree is that of the underlying lambda-term.

Moreover a !-derivation $\tilde{\mathcal{D}}$ can also be translated into a syntactic tree with boxes (forgetting about types): a box is put around each argument of a $(\Rightarrow e)$ application.

Observe that if a canonical abstract derivation $Q$ and a !-derivation $\tilde{\mathcal{D}}$ correspond to the same term $t$, then their associated trees can only differ by the boxes.

We say a canonical simple form $Q$, with tree $\mathcal{T}_1$, is *compatible* with the !-derivation $\tilde{\mathcal{D}}$, with tree $\mathcal{T}_2$, if the following conditions hold:

- any box of $\mathcal{T}_2$ corresponds to a box of $\mathcal{T}_1$ (that is to say $\mathcal{T}_1$ is obtained from $\mathcal{T}_2$ by adding some boxes);

- for any box $\mathcal{B}_2$ of $\mathcal{T}_2$ with input variable $x$ (that is to say $x$ is a free variable of the corresponding term) then: any box $\mathcal{B}_1$ of $\mathcal{T}_1$ containg $\mathcal{B}_2$ also contains the $\lambda$ node abstracting $x$ (and no such box exists if $x$ is not abstracted).

Graphically the second condition amounts to say that in $\mathcal{T}_1$ no box can be closed below $\mathcal{B}_2$ and have $x$ as input. These two conditions can be checked by one traversal of both trees, and by comparing the tree of each element of the finite set $C(t)$ to that of $\tilde{\mathcal{D}}$ we can determine $\tilde{C}(t)$ and thus complete step b).

We now consider step c). Let $Q$ be an element of $\tilde{C}(t)$ and $PT(Q) = <\theta, \sigma, \mathcal{C}>$. The procedure $PT(.)$ assignes to each box of (the tree associated to) $Q$ a distinct parameter $n$. Let us denote by $\mathcal{B}_1, \ldots, \mathcal{B}_k$ the boxes of $Q$ corresponding to boxes of $\tilde{\mathcal{D}}$ and by $n_1, \ldots n_k$ the corresponding parameters assigned by $PT(.)$.

From the results on $PT(.)$ we know that any solution $X$ of $\tilde{\mathcal{C}}$ induces an EAL derivation for $t$. It can also define an LAL derivation in the following way: each box $\mathcal{B}_i$ ($1 \leq i \leq k$) is instantiated into one !-box and ($X(n_i)$-1) $\S$-boxes (so possibly 0); all other boxes are instantiated by $\S$-boxes. For each $\S$-box (($\S$ i) rules) the type (! or $\S$) of the discharged variables can be chosen so as to get a valid derivation. Finally an LAL type derivation for $t$ obtained in this way is the translation by $(.)^*$ of a DLAL derivation.

**Remark 32** *This procedure is not very satisfactory because it starts by determining a distribution of !-boxes (with several possibilities) and then enumerates all canonical simple derivations before searching which ones match the distribution of !-boxes. It would be more efficient to compute directly the canonical simple derivations corresponding to the distribution of !-boxes.*